\documentclass[sigconf]{acmart}

\AtBeginDocument{%
  }
\usepackage{algorithm}
\usepackage{algpseudocode}
\usepackage{amsmath}
\usepackage{tikz}
\usetikzlibrary{arrows.meta, positioning}
\usepackage{alltt}
\usepackage{makecell} 
\usepackage{booktabs}
\usepackage{tcolorbox}
\usepackage{placeins}
\usepackage{float}
\usepackage{graphicx}  
\usepackage{svg}        
\usepackage{dblfloatfix}  
\usepackage{tikz}
\usetikzlibrary{arrows.meta, positioning, calc}

\usepackage{caption}
\setcopyright{none}
\renewcommand\footnotetextcopyrightpermission[1]{}

\begin{document}

\title{From Textual Requirements to Microservice Architectures: A Comprehensive Evaluation of LLM-Based Design Synthesis}

\author{Danyllo Albuquerque}
\affiliation{%
  \institution{VIRTUS/UFCG \\Federal University of Campina Grande (UFCG), Paraiba}
  \country{Brazil}
  }
\email{danyllo@copin.ufcg.edu.br}

\author{José Renan}
\affiliation{%
  \institution{Federal University of Campina Grande (UFCG), Paraiba}
  \country{Brazil}
  }
\email{jose.pereira@embedded.ufcg.edu.br}

\author{Guillermo Rodríguez}
\affiliation{%
  \institution{Consejo Nacional de Investigaciones Científicas y Técnicas (CONICET), Universidad Nacional del Centro de la Provincia de Buenos Aires (UNICEN), Argentina}
  \country{Argentina}
  }
\email{exarodriguez@gmail.com}

\author{Jorge Andrés Díaz-Pace}
\affiliation{%
  \institution{Consejo Nacional de Investigaciones Científicas y Técnicas (CONICET), Universidad Nacional del Centro de la Provincia de Buenos Aires (UNICEN)}
  \country{Argentina}
  }
\email{adiazpace@gmail.com}

\author{Emanuel Dantas}
\affiliation{%
  \institution{VIRTUS/UFCG \\Federal Institute of Pernambuco (IFPE), Brazil}
  \country{Brazil}
  }
\email{emanuel.dantas@virtus.ufcg.edu.br}

\author{Ademar França}
\affiliation{%
  \institution{Rui Barbosa State School, Rio Grande do Norte, Brazil}
  \country{Brazil}
  }
\email{ademar.sousa@virtus.ufcg.edu.br}

\author{Mirko Perkusich}
\affiliation{%
  \institution{VIRTUS/UFCG \\Federal University of Campina Grande (UFCG), Paraiba, Brazil}
  \country{Brazil}
  }
\email{mirko@virtus.ufcg.edu.br}

\author{Kyller Gorgônio}
\affiliation{%
  \institution{VIRTUS/UFCG \\Federal University of Campina Grande (UFCG), Paraiba, Brazil}
  \country{Brazil}
  }
\email{kyller@virtus.ufcg.edu.br}

\author{Angelo Perkusich}
\affiliation{%
  \institution{VIRTUS/UFCG \\Federal University of Campina Grande (UFCG), Paraiba, Brazil}
  \country{Brazil}
  }
\email{perkusic@virtus.ufcg.edu.br}

\renewcommand{\shorttitle}{From Textual Requirements to Microservice Architectures}

\renewcommand\footnotetextcopyrightpermission[1]{}

\setcopyright{none}
\settopmatter{printacmref=false}
\renewcommand{\shortauthors}{Silva et al.}



\begin{abstract}
[\textbf{Context}]. Microservice architectures have become a dominant paradigm for modernizing monolithic systems. However, identifying appropriate services remains a challenging and largely manual task. Existing decomposition approaches are predominantly code-centric, limiting their applicability in early design stages where only textual requirements are available. [\textbf{Problem}]. Despite recent advances in Large Language Models (LLMs), there is still limited empirical evidence regarding their ability to synthesize complete microservice architectures directly from natural-language requirements, including both service definitions and inter-service interactions. [\textbf{Goal}]. This study investigates whether an LLM can bridge the gap between requirements engineering and architectural design by generating microservice architectures solely from textual requirements, and evaluates the structural agreement and perceived quality of the generated solutions. [\textbf{Method}]. We conduct a mixed-method study using the OpenAI o3 model under zero-shot (ZS) and few-shot (FS) prompting strategies across two systems (Bookstore and PetClinic), with one execution per system and prompting condition. Generated architectures are evaluated through (i) quantitative comparison with implemented reference architectures using precision, recall, and F1-score for service identification and communication recovery, and (ii) a blinded expert assessment of correctness, completeness, modularity, and plausibility, supplemented by a structured descriptive synthesis of open-ended feedback. [\textbf{Results}]. The findings indicate that OpenAI o3 can identify services from requirements, achieving higher agreement with the references under FS prompting (F1 $\approx$ 0.79 for ZS and $\approx$ 0.97 for FS). Communication recovery is more challenging, with ZS producing overly dense architectures characterized by high recall but low precision (F1 $\approx$ 0.61). FS prompting improves structural agreement and perceived architectural quality, achieving F1 $\approx$ 0.82 for communication recovery while reducing unsupported dependencies. Expert evaluation corroborates these findings, indicating that FS-generated architectures are consistently perceived as more modular, coherent, and plausible than ZS outputs. [\textbf{Conclusion}]. Within the evaluated scope, OpenAI o3 shows potential to support requirements-driven architectural synthesis, particularly when guided by minimal exemplar-based prompting. The results should be interpreted as model- and context-specific evidence from two relatively small systems and a single execution per condition, rather than as model-independent proof of effectiveness.
\end{abstract}



\sloppy
\maketitle

\section{Introduction}
\label{sec:introduction}

Legacy enterprise systems continue to support a substantial portion of organizational operations, yet many of them were originally designed as monolithic applications. While this architectural style simplifies early development, it increasingly conflicts with modern requirements such as scalability, continuous delivery, and rapid adaptation to evolving business needs ~\cite{seacord2003modernizing, kaloudis2024evolving}. As a result, organizations are actively pursuing modernization strategies to transition toward more modular and flexible architectures ~\cite{taibi2017decomposition,ponce2019systematic}.

Among these strategies, the migration to microservice-based architectures has emerged as a dominant paradigm. Microservices decompose applications into loosely coupled, independently deployable services aligned with business capabilities~\cite{romani2022towards, saucedo2025migration}. This paradigm enables scalability, resilience, and faster evolution~\cite{newman2015building,dragoni2017microservices}. However, these benefits critically depend on the identification of appropriate service boundaries and the definition of coherent and well-structured inter-service interactions.

Determining such boundaries and interactions remains a challenging and largely manual task. Existing approaches to microservice decomposition are predominantly code-centric, relying on static dependencies, runtime traces, or repository mining~\cite{abgaz2023decomposition}. While effective in reverse-engineering existing systems, these approaches often fail to capture the semantic intent embedded in requirements and domain descriptions. As a result, they may produce architectures that are structurally consistent at the code level but misaligned with business logic, both in terms of service responsibilities and interaction patterns~\cite{nogueira2024insights}.

This limitation exposes a fundamental gap: architectural decomposition is inherently a requirements-driven problem, yet current automated approaches largely ignore requirements as primary input. In practice, architects still rely on manual interpretation of textual artifacts (e.g., user stories and business rules) to derive both service boundaries and inter-service interactions, making the process time-consuming, subjective, and difficult to scale~\cite{capilla201610}.

In this context, recent advances in Large Language Models (LLMs) offer a promising alternative ~\cite{chen2021codex,nijkamp2022codegen,wang2023llmse}. LLMs demonstrate strong capabilities in natural language understanding, abstraction, and structured generation, enabling them to extract domain concepts directly from textual descriptions ~\cite{hou2024large, zheng2025towards}.  

While prior work has explored LLMs in software engineering tasks such as code generation and requirements analysis ~\cite{albuquerque2022managing}, their role in end-to-end architectural synthesis from requirements remains largely unexplored and under-evaluated. This raises a key question: \textit{Can LLMs bridge the gap between textual requirements and architectural design by synthesizing coherent microservice architectures directly from natural-language inputs?}

To address this question, we conduct an empirical study of OpenAI o3's ability to synthesize microservice architectures directly from textual requirements. Specifically, we evaluate the structural agreement and perceived architectural quality of architectures generated under zero-shot (ZS) and few-shot (FS) prompting strategies. The study combines quantitative comparison with implemented reference decompositions~\cite{imranur2019curated}, blinded expert ratings, and a structured descriptive synthesis of open-ended expert feedback. Because the experiment covers one model, two relatively small systems, and one execution per prompting condition, the resulting conclusions are intentionally scoped to this empirical setting.

\sloppy
This article makes the following contributions:

\begin{itemize}
    \item An empirical evaluation of LLMs for generating microservice architectures directly from textual requirements, without relying on code or runtime artifacts;

    \item A mixed-method evaluation framework that combines quantitative structural agreement (service identification and inter-service interactions) (RQ1--RQ2), blinded expert ratings of architectural quality (RQ3), and a structured descriptive synthesis of open-ended feedback (RQ4);

    \item An analysis of how ZS and FS prompting strategies influence structural outcomes, revealing their impact on service identification and interaction recovery accuracy (RQ1–RQ2);

    \item An examination of expert ratings and open-ended feedback that operationalizes correctness, completeness, modularity, and plausibility, while preserving divergent assessments and characterizing recurring strengths, weaknesses, and design issues in the generated architectures (RQ3--RQ4);

    \item Evidence that minimal exemplar-based prompting improves both structural alignment and perceived architectural quality, highlighting its role as an effective mechanism for requirements-driven design support.
\end{itemize}

These contributions provide empirical evidence that LLMs can support architectural reasoning in scenarios where design decisions must be derived from high-level textual specifications, contributing to ongoing efforts toward AI-assisted software modernization and architectural evolution. Building on our previous study~\cite{sbcars}, which provided an initial exploratory investigation, this work advances the state of the art by introducing a rigorous empirical evaluation of LLM-based architectural synthesis, including a refined methodological design, an additional research question (RQ4), and a substantially expanded analysis combining quantitative metrics, qualitative evidence, and stronger theoretical grounding and implications.

The remainder of this article is organized as follows: Section~\ref{sec:background} outlines the conceptual and architectural foundations underlying this study. Section~\ref{sec:rel_work} reviews related work. Section~\ref{sec:methodology} describes the study design and evaluation procedure. Section~\ref{sec:results} reports the empirical results. Section~\ref{sec:analysis} provides a deeper interpretation of the findings. Section~\ref{sec:implications} discusses implications for research and practice. Section~\ref{sec:threats} outlines threats to validity. Finally, Section~\ref{sec:final_remarks} concludes the article.

\section{Foundations}
\label{sec:background}

This section presents the conceptual and architectural foundations underlying this study. It introduces the principles of microservice decomposition, the challenges of defining service boundaries and inter-service interactions, and the role of textual requirements in early-stage architectural design~\cite{ferrari2017nlp}.

Monolithic architectures consolidate application components, data-access logic, and deployment into a single codebase and execution unit. While suitable in early development stages, this architectural style becomes increasingly problematic as systems evolve and accumulate tightly coupled components, cross-cutting dependencies, and implicit business rules \cite{abgaz2023decomposition, hassan2024migrating}. Under these conditions, even localized changes may require rebuilding, retesting, and redeploying the entire application, increasing delivery time and operational risk.

Microservice architectures address these limitations by decomposing applications into smaller, autonomous, and independent services, each aligned with a specific business capability or bounded responsibility~\cite{newman2015building, dragoni2017microservices}. In this paradigm, services encapsulate domain logic and expose functionality through well-defined interfaces, while the overall system behavior emerges from the inter-service interactions, realized through lightweight communication mechanisms such as REST APIs or messaging \cite{wolff2016microservices}. This paradigm improves modular evolution, supports technological heterogeneity, enhances fault isolation, and enables fine-grained scalability \cite{hassan2024migrating}. However, these benefits depend on the quality of the resulting decomposition in terms of high internal cohesion and low external coupling within services, and the correctness and efficiency of inter-service interactions~\cite{tapia2023research, Vera-Rivera2021Defining}.

A central challenge in this transition is the accurate identification of service boundaries and the inter-service interactions. In legacy systems, such boundaries and interactions are rarely explicit, as responsibilities are often dispersed across modules and shaped by incremental changes over time~\cite{bushong2021microservice}. Consequently, microservice decomposition is not merely a technical refactoring task, but an architectural recovery process that requires reconstructing domain structure and mapping it to modular service designs, as well as defining how these services coordinate and communicate~\cite{joselyne2021systematic}.

Poorly defined boundaries may lead to overly coarse services that replicate monolithic limitations, or excessively fine-grained services that increase communication overhead and fragment business logic \cite{saucedo2024variability}. Similarly, poorly designed interactions may result in tightly coupled communication patterns, redundant dependencies, or inefficient data flows, ultimately degrading system modularity and performance~\cite{oumoussa2024evolution}.

To mitigate migration risks, incremental modernization strategies such as the \emph{Strangler Fig} pattern are commonly adopted, allowing legacy functionality to be gradually replaced by new services \cite{ponnusamy2023navigating}. Additionally, \emph{Domain-Driven Design} (DDD) is frequently used to align decomposition with bounded contexts, domain concepts, and business rules \cite{sistla2023domain}. This perspective emphasizes that effective microservice design depends on structural modularization and preserving the semantic integrity of the problem domain \cite{velepucha2023survey}.

Given the complexity of manual migration, a wide range of automated and semi-automated approaches have been proposed to support service identification \cite{abgaz2023decomposition}. These approaches span structural, behavioral, and semantic techniques \cite{ponce2019systematic}. Structural approaches rely on static analysis of code elements and dependencies \cite{fritzsch2018monolith, abgaz2023decomposition, oumoussa2024evolution}, while search-based and evolutionary techniques explore alternative decompositions using optimization strategies \cite{gysel2016service, mazlami2017extraction}. Other methods leverage repository mining and co-change analysis to uncover latent service boundaries grounded in maintenance behavior \cite{assunccaocontemporary}. Behavioral and hybrid approaches incorporate runtime information such as execution traces and logs to capture dynamic interactions \cite{joselyne2021systematic}. Complementarily, semantic and lexical techniques exploit domain vocabulary, documentation, and embeddings to identify conceptually coherent modules \cite{saucedo2024variability, saucedo2024migration, alsayed2024microdec, sellami2025contrastive}.

Most existing approaches rely on code-level artifacts (e.g., source code or execution traces) to support decomposition. These methods assume that such artifacts are available and sufficiently informative to guide the identification of service boundaries and interactions. While effective in scenarios where implementation data is accessible, this reliance limits their applicability in early design stages, greenfield contexts, or requirement-driven settings where such artifacts are not yet available~\cite{joselyne2021systematic, saucedo2024variability}. Furthermore, these approaches may yield decompositions that are structurally consistent but only partially aligned with domain concepts, stakeholder intent, or textual requirement descriptions.

Assessing the quality of a candidate decomposition is itself a non-trivial problem. In the literature, microservice quality is commonly evaluated through architectural properties such as cohesion, coupling, granularity, and conceptual consistency \cite{tapia2023research}. Representative metrics include Structural Modularity Quality (SMQ) and Conceptual Modularity Quality (CMQ), which respectively assess structural separation and conceptual coherence \cite{Vera-Rivera2021Defining}, as well as semantic cohesion measures \cite{Mohottige2025Reengineering}, and granularity indicators \cite{Weerasinghe2026From}. However, these metrics do not fully eliminate the need for human judgment, as architectural plausibility also depends on domain alignment and operational feasibility.

The emergence of Large Language Models (LLMs), such as GPT-4, Gemini, Llama, and DeepSeek, suggests a new paradigm for architecture-oriented automation \cite{hou2024large, dhar2024can, he2024llm, ataei2025elicitron}. Unlike traditional code-centric approaches, LLMs can process natural-language artifacts (e.g., requirements, user stories, and domain descriptions) and extract latent domain concepts, relationships, and abstractions \cite{zhao2023llmre,wang2023llmse}. This capability enables them to potentially generate structured outputs, including service candidates and interaction patterns, directly from textual inputs \cite{alsayed2024microdec, sellami2025contrastive, trabelsi2025systematic}.

By operating at the domain-semantic level rather than implementation details, LLMs can support architectural reasoning in early design stages, when code-level artifacts are not yet available. This shift indicates that LLMs may act as a bridge between business-level descriptions and architectural design decisions, opening new possibilities for requirements-driven microservice decomposition.

\section{Related Work}
\label{sec:rel_work}

Prior research on microservice decomposition has predominantly focused on implementation-driven strategies. Most approaches infer service candidates (and inter-service interaction) from source code structure, execution traces, repository history, or dependency graphs, using clustering, heuristic rules, or search-based optimization to identify modular partitions \cite{abgaz2023decomposition, fritzsch2018monolith, gysel2016service, mazlami2017extraction, assunccaocontemporary, joselyne2021systematic}. These methods have proven effective in modernization scenarios, where code-level artifacts are readily available, but they generally assume that such artifacts are sufficiently rich and reliable to guide decomposition decisions.

Building on this foundation, a second line of work incorporates semantic, lexical, API-level, and learning-based signals to improve service identification \cite{saucedo2024variability, baresi2017microservices, al2020extracting, Jin2019Service, saucedo2024migration, alsayed2024microdec, sellami2025contrastive, trabelsi2025systematic}. By leveraging domain vocabulary, interface descriptions, and embedding-based representations, these approaches aim to better capture conceptual coherence and business functionality. Nevertheless, despite this shift toward richer representations, they still predominantly rely on code-level artifacts (e.g., code, APIs, execution data, or hybrid inputs) rather than treating textual requirements as a primary source of architectural knowledge.

More recently, research in software engineering has explored the use of LLMs and related AI techniques for design-oriented tasks such as requirements analysis, architectural decision support, and model-driven generation \cite{hou2024large, dhar2024can, he2024llm, ataei2025elicitron}. These studies demonstrate that language-based models can support reasoning over textual artifacts and help derive design-relevant information \cite{chen2021codex,austin2021program,wang2023llmse,ferrari2017nlp,zhao2023llmre}. However, most of this work focuses on isolated activities or intermediate artifacts rather than directly addressing the end-to-end synthesis of microservice architectures from requirements.

Among the studies closest to our research objective, Abgaz et al.~\cite{abgaz2023decomposition} provide an important baseline by highlighting that decomposition research remains largely centered on implementation artifacts. Moving toward stronger semantic support, MicroDec~\cite{alsayed2024microdec} and MonoEmbed~\cite{sellami2025contrastive} integrate structural analysis with embeddings and learning-based techniques to identify service candidates, yet still operate primarily over code-level inputs. In a complementary direction, Dhar et al.~\cite{dhar2024can} investigate LLM-based generation of architectural design decisions, while Ataei et al.~\cite{ataei2025elicitron} explore LLM-driven requirements elicitation; both contributions are relevant to early-stage design, but focus on adjacent artifacts rather than on the synthesis and evaluation of complete microservice architectures.

\textbf{Research Gap}. Despite these advances, a fundamental gap remains at the intersection of requirements engineering and architectural design. Existing approaches predominantly rely on implementation artifacts or leverage LLMs for isolated or intermediate design tasks, but do not systematically evaluate their ability to synthesize complete microservice architectures directly from textual requirements. As a result, there is still limited empirical understanding of how accurately such architectures reflect reference decompositions, how well they capture inter-service relationships, and to what extent they are perceived as coherent and plausible by experts.

To address this gap, this study adopts a requirements-driven perspective, treating textual specifications as the sole input to generate complete architectural artifacts—including services and inter-service interactions—and evaluating them through a combined quantitative and expert-based framework. In doing so, it bridges the gap between requirements engineering and architectural design, providing systematic empirical evidence on the feasibility and limitations of LLM-based architectural synthesis. These limitations motivate the empirical investigation presented in this study.

\section{Study Settings}
\label{sec:methodology}

This section presents the study goals and research questions (Section~\ref{subsec:rqs}), followed by the research design, which outlines the methodological steps adopted in this study (Section~\ref{subsec:research_steps}).

\subsection{Goals and Research Questions}
\label{subsec:rqs}

The goal of this study is to assess whether LLMs can generate microservice architectures directly from natural-language requirements, without relying on source-code artifacts. To address this objective, we adopt a dual evaluation perspective. First, we analyze the structural correspondence between LLM-generated and reference architectures, focusing on service identification and inter-service interactions. Second, we assess the \textit{perceived architectural quality} of the generated outputs through expert evaluation, considering correctness, completeness, modularity, and plausibility. Together, these perspectives provide a comprehensive view that combines objective alignment with practitioner-oriented judgment. This evaluation is operationalized through a set of research questions (RQs) that capture these complementary aspects and structure the analysis across both quantitative and qualitative dimensions. Table~\ref{tab:rqs} summarizes the RQs along with their descriptions and motivations.

\begin{table*}[!ht]
\centering
\small
\caption{Research questions, descriptions, and motivations}
\label{tab:rqs}
\resizebox{0.95\linewidth}{!}{
\begin{tabular}{p{1.0cm} p{6.05cm} p{9.5cm}}
\toprule
\textbf{RQ} & \textbf{Description} & \textbf{Motivation} \\
\midrule

\textbf{RQ1} & 
Can LLMs identify the individual microservice components from textual requirement descriptions? 
& 
Focuses on decomposition capability, i.e., whether an LLM can transform natural-language requirements into a plausible set of service components. This is fundamental, as microservice design begins with identifying bounded responsibilities and coherent service candidates. \\

\midrule

\textbf{RQ2} & 
How accurately do LLMs recover the inter-service communication links present in the reference architecture? 
& 
Addresses architectural fidelity by evaluating whether the model can reconstruct services and their communication relationships, which are essential for defining an operational architecture. \\

\midrule

\textbf{RQ3} & 
How do experts evaluate the quality of LLM-generated architectures in terms of correctness, completeness, modularity, and plausibility? 
& 
Introduces a practitioner-oriented perspective, assessing whether generated architectures are perceived as structurally sound and realistic by experienced software architects. \\

\midrule

\textbf{RQ4} & 
What strengths, weaknesses, and design patterns do experts identify in LLM-generated architectures based on qualitative feedback? 
& 
Extends the evaluation to qualitative insights, capturing recurring strengths, weaknesses, and design patterns that help explain the behavior of LLM-generated architectures beyond numerical metrics. \\

\bottomrule
\end{tabular}}
\end{table*}

Together, these RQs define a cohesive analytical framework that links structural recovery, architectural fidelity, and expert-based evaluation, enabling a systematic assessment of both the structural and practical validity of the generated architectures.

\subsection{Research Design}
\label{subsec:research_steps}

\begin{figure*}[!ht]
\centering
\resizebox{0.95\textwidth}{!}{
\begin{tikzpicture}[
    font=\small,
    >=Stealth,
    node distance=0.9cm and 0.6cm,
    stage/.style={
        rectangle,
        rounded corners=4pt,
        draw=#1!70!black,
        very thick,
        fill=#1!8,
        text width=4.2cm,
        minimum height=7.2cm,
        align=left
    },
    titlebox/.style={
        rectangle,
        rounded corners=3pt,
        draw=#1!70!black,
        fill=#1!20,
        text centered,
        font=\bfseries,
        minimum height=0.95cm,
        text width=3.6cm
    },
    io/.style={
        font=\scriptsize,
        align=left,
        text width=3.7cm
    },
    flow/.style={
        -{Stealth[length=3mm,width=2mm]},
        thick,
        draw=black!70
    }
]

\node[stage=blue] (s1) {
    \textbf{\textcolor{red!75!black}{Input:}} textual descriptions of system requirements and reference architectural artifacts.\vspace{0.25cm}

    \textbf{\textcolor{green!40!black}{Activities:}}
    \begin{itemize}
        \item selection of the case studies (Bookstore and PetClinic);
        \item curation and standardization of requirements;
        \item extraction of the reference architecture from the systems;
        \item normalization of artifacts for comparison.
    \end{itemize}

    \textbf{\textcolor{blue!70!black}{Output:}} 
      \begin{itemize}
        \item consolidated requirements; 
        \item standardized reference architectures.
      \end{itemize}

};
\node[titlebox=blue, above=0.15cm of s1.north] (t1) {1. Dataset Preparation};

\node[stage=orange, right=0.9cm of s1] (s2) {
    \textbf{\textcolor{red!75!black}{Input:}} standardized requirements and architectural generation instructions.\vspace{0.25cm}

    \textbf{\textcolor{green!40!black}{Activities:}}
    \begin{itemize}
        \item LLM selection;
        \item definition of the expected output format;
        \item application of \textit{zero-shot};
        \item application of \textit{few-shot};
        \item generation of candidate architectures with services, responsibilities, and interactions.
    \end{itemize}

    \textbf{\textcolor{blue!70!black}{Output:}} 
     \begin{itemize}
        \item candidate architectures generated by the LLM under the ZS and FS conditions.
        
     \end{itemize}

};
\node[titlebox=orange, above=0.15cm of s2.north] (t2) {2. Prompting-Based Generation};

\node[stage=teal, right=0.9cm of s2] (s3) {
    \textbf{\textcolor{red!75!black}{Input:}} architectures generated by the LLM and reference architectures.\vspace{0.25cm}

    \textbf{\textcolor{green!40!black}{Activities:}}
    \begin{itemize}
        \item comparison of the identified services;
        \item comparison of interactions between services;
        \item classification into correct, missing, and extra elements;
        \item calculation of precision, recall, and F1-score.
    \end{itemize}

    \textbf{\textcolor{blue!70!black}{Output:}} 
     \begin{itemize}
        \item quantitative metrics;
        \item characterization of architectural errors.
    \end{itemize}

};
\node[titlebox=teal, above=0.15cm of s3.north] (t3) {3. Quantitative Evaluation};

\node[stage=purple, right=0.9cm of s3] (s4) {
    \textbf{\textcolor{red!75!black}{Input:}} requirements, anonymized diagrams/descriptions, and evaluation form.\vspace{0.25cm}

    \textbf{\textcolor{green!40!black}{Activities:}}
    \begin{itemize}
        \item Preparation of evaluation materials and recruitment of expert participants;
        \item quantitative assessment of architectures;
        \item qualitative assessment of architectures.
    \end{itemize}

    \textbf{\textcolor{blue!70!black}{Output:}} 
    \begin{itemize}
        \item scores for correctness, completeness, modularity, and plausibility;
        \item qualitative feedback regarding strengths, risks, and design choices.
    \end{itemize}

};
\node[titlebox=purple, above=0.15cm of s4.north] (t4) {4. Expert Validation};

\draw[flow] (s1.east) -- (s2.west);
\draw[flow] (s2.east) -- (s3.west);
\draw[flow] (s3.east) -- (s4.west);

\draw[dashed, gray!60] 
    ($(s1.south)+(0,-0.35)$) -- ($(s4.south)+(0,-0.35)$);

\node[font=\bfseries\small, align=center] at ($(s1.south)!0.5!(s4.south)+(0,-0.8)$)
{};

\end{tikzpicture}
}
\caption{Methodological workflow adopted in this study.}
\label{fig:methodology_workflow}
\end{figure*}
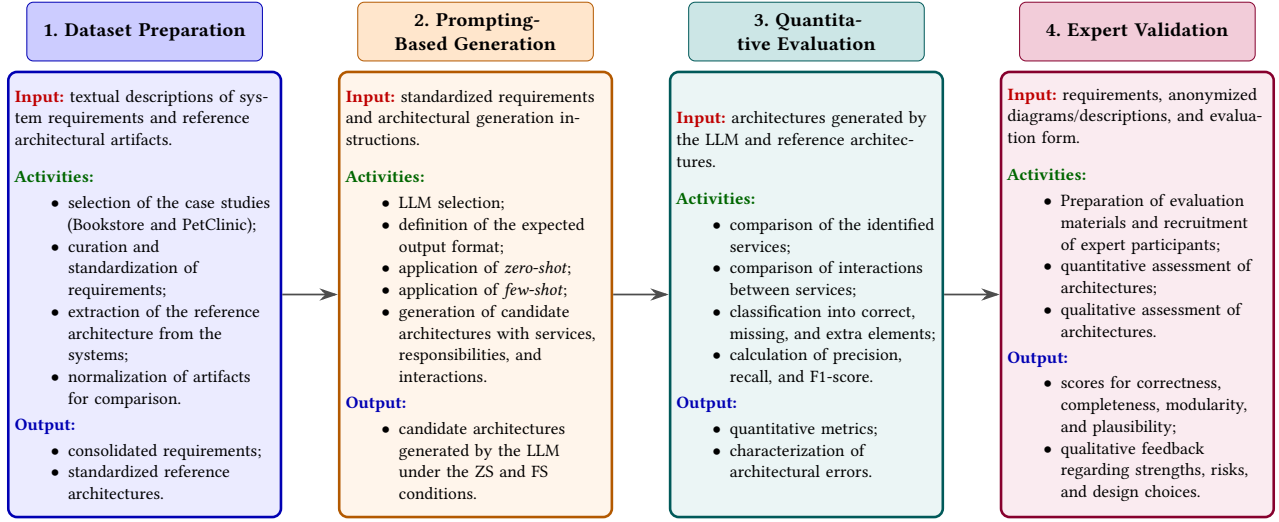

To answer the RQs, we designed a four-step methodology that integrates artifact preparation, prompt-based architectural generation, quantitative comparison with reference systems, and expert-based validation. Figure~\ref{fig:methodology_workflow} presents an overview of the workflow and its corresponding inputs, activities, and outputs.

The methodological design combines quantitative and qualitative evaluation. The quantitative component measures the degree of overlap between LLM-generated and reference architectures for service identification and inter-service communication, directly addressing RQ1 and RQ2. The qualitative component relies on expert judgment to assess how understandable, plausible, and architecturally sound the generated outputs appear in practice, thereby addressing RQ3. At the same time, the open-ended feedback supports the identification of recurring strengths, weaknesses, and design patterns for RQ4. This mixed-method design was chosen because purely structural similarity does not fully capture architectural quality. An architecture may resemble a numerically decomposed reference while still being perceived as poorly modularized or operationally implausible. Conversely, a design may differ from the implementation yet still be judged reasonable from a business or design perspective.

The remainder of this section details each step of the methodology. All artifacts used in the study are publicly available to support transparency and reproducibility (see Artifacts Availability Section).

\subsubsection*{Step 1: Dataset Preparation}
\label{subsec:dataset_preparation}

This step establishes the foundational artifacts required for the study, including selecting case studies, preparing requirement descriptions, and extracting reference architectures. Its objective is to ensure that both the LLM's input and the baseline for comparison are consistent, representative, and suitable for systematic evaluation.

\textbf{Selection of case studies}. This study employed two software systems from the dataset curated by Imranur et al.~\cite{imranur2019curated}: (i) \textit{Bookstore} and (ii) \textit{PetClinic}. Both systems provide accessible requirement descriptions and sufficiently explicit architectural artifacts for systematic comparison, and they represent distinct functional domains. Their limited scale also makes it feasible to inspect services and interactions consistently during the quantitative comparison and expert review.

The two cases were selected for controlled, in-depth experimentation, not as a statistically representative sample of microservice systems. Both contain a relatively small number of services and comparatively well-defined requirements. Consequently, they provide evidence about the feasibility of the evaluated procedure in bounded settings, but do not establish that the same performance will hold for larger systems, heterogeneous portfolios, safety-critical applications, or architectures with substantially denser interaction topologies. Replication with additional systems is required to assess external validity.

\textbf{Requirements curation}. To support requirement-driven architectural generation, the natural-language requirements of each system were curated and standardized before being presented to the model. This curation process had two main objectives: improving clarity and preserving semantic fidelity. In particular, the requirements were revised to remove unnecessary redundancy, reduce ambiguity, and normalize domain terminology while preserving the original functionality and business intent of the system. The curation focused on textual clarification rather than functional modification so that the evaluation would reflect the model’s architectural reasoning rather than differences in requirement phrasing.

Each system was ultimately represented by a consolidated set of at least ten functional requirements expressed in declarative form. The requirements were organized to describe the target system's expected functionalities in a coherent, self-contained manner. This preparation step was necessary because LLM-based generation is sensitive to ambiguity, lexical inconsistency, and incomplete contextualization. By standardizing the phrasing of requirements, we sought to reduce noise unrelated to the architectural reasoning capability being evaluated.

Importantly, this curation was not intended to redesign or enrich the systems with additional features. Rather, it aimed to present the original business functionality in a clearer, more consistent textual form that could be consistently interpreted by both the model and the human evaluators.

\textbf{Extraction of the reference architecture}. We reused the service and dependency information distributed with the curated dataset~\cite{imranur2019curated}, which derives architectural characteristics from the corresponding open-source implementations. The dataset's analysis pipeline uses \textit{SLOCcount} to characterize source-code size and \textit{MicroDepGraph} to recover and visualize inter-service dependencies. MicroDepGraph analyzes service declarations in Docker Compose files and internal API calls in Java source code to construct the dependency graph. In this study, the resulting service and dependency records for Bookstore and PetClinic were the starting point for the reference architectures.

We normalized this information into a common representation that documented: (i) the list of implemented services, (ii) the main responsibility assigned to each service, and (iii) the directed communication relationships among services. The same abstraction level was used for the reference, ZS, and FS alternatives so that formatting and notation would not determine the comparison or the expert judgments.

These references are realistic because they originate from existing implementations, but their quality and representativeness are bounded by the selected repositories and by the extraction procedure. They are therefore treated as implemented baselines, not as optimal or unique decompositions. Precision, recall, and F1-score measure agreement with these baselines. An element classified as \textit{extra} can still constitute a defensible alternative design choice; the label only indicates that the element is absent from the implemented reference. The blinded expert assessment complements this structural comparison by evaluating whether such alternatives remain correct, complete, modular, and plausible.

\textbf{Artifact normalization}. To support consistent downstream evaluation, all artifacts were standardized before use. Requirements were formatted uniformly across systems, and reference architectures were rendered as structured diagrams and textual descriptions using the same level of abstraction later used for the LLM-generated outputs. This normalization step was especially important for the expert-review stage, as it reduced the chance that presentation format, notation style, or diagram layout would bias evaluators toward one alternative.

At the end of this step, we obtain a consistent set of requirements and reference architectures that serve as input to the LLM and as a baseline for quantitative comparison. These artifacts are then used in the next step to generate candidate architectures under different prompting strategies.

\subsubsection*{Step 2: Prompt-Based Generation}
\label{subsec:prompt_generation}

This step focuses on generating candidate architectural decompositions from the curated requirements using LLM-based prompting strategies. Its objective is to assess how different prompting conditions influence the model’s ability to infer service boundaries, responsibilities, and inter-service interactions.

\textbf{LLM selection}. To generate architectural decompositions from natural-language requirements, we used the OpenAI o3 model via the OpenAI API.\footnote{https://platform.openai.com/docs/models/o3} The model was selected because the task requires natural-language understanding, abstraction over domain concepts, structured output synthesis, and multi-step reasoning about decomposition and dependencies. However, the experiment was not designed as a comparison among model families. Accordingly, the empirical results directly characterize OpenAI o3 under the reported prompts and execution period; broader claims about LLM-based architectural synthesis remain hypotheses to be tested through cross-model replication.

\textbf{Definition of output format}. The prompts were designed to elicit a structured architectural proposal rather than a free-form narrative. For each system, the model was instructed to read the complete set of functional requirements and produce:

\begin{itemize}
    \item a set of proposed microservices;
    \item a concise name for each service;
    \item a description of the main responsibilities of each service; and
    \item an explicit indication of which services communicate with which others.
\end{itemize}

This output structure was chosen because it aligns directly with the study objectives. The identification of services addresses RQ1, while the explicit communication links address RQ2. Requiring service names and responsibilities also improves interpretability and supports the later expert assessment for RQ3.

To reduce uncontrolled variability, the prompts emphasized that the model should propose services grounded in the given requirements and avoid excessive speculation beyond the textual evidence. Even so, the prompts did not specify the number of services or any target decomposition template, since doing so would constrain the model too strongly and interfere with the evaluation's purpose.

\textbf{Application of prompting strategies}. To analyze the effect of prior architectural guidance, we adopted two prompting strategies:

\begin{itemize}
    \item \textit{ZS prompting}: the model received only the task instructions and the complete requirements of the target system. No example of architectural decomposition was provided. This condition evaluates the model’s unaided ability to infer service boundaries and interactions directly from the requirements.
    
    \item \textit{FS prompting}: the model received the same task instructions and requirements, but the prompt additionally included one illustrative example of a different system decomposition. This example was formatted as a structured list of services, responsibilities, and interactions. The purpose of this condition was to determine whether minimal exposure to the expected output format and reasoning style would improve the coherence and alignment of the generated architecture.
\end{itemize}

The authors manually curated the FS example from a system different from Bookstore and PetClinic. The exemplar did not reproduce the target requirement text and was intended to demonstrate the expected representation---a structured mapping from requirements to services, responsibilities, and directed interactions---at a comparable abstraction level. However, the original study did not quantify the exemplar's domain similarity or decomposition-pattern similarity to the target systems. The exemplar may therefore still act as a structural prior by suggesting a particular granularity or interaction density. Consequently, the observed FS improvement cannot be attributed exclusively to output-format clarification. A richer instruction-only prompt could potentially provide part of the same guidance without an explicit decomposition example, and a direct comparison between exemplar-based and instruction-only guidance remains necessary.

\textbf{Prompt design considerations}. All prompts were written to be instruction-focused and as neutral as possible. We avoided wording that explicitly suggested particular services, architectural patterns, or implementation technologies for the target systems. This was important to reduce the risk that the prompt itself would impose a decomposition. Likewise, the two prompting conditions differed only in the presence or absence of the illustrative example, preserving comparability across the two settings.

A complete set of prompts is available in the supplementary repository to support replication and secondary analysis.

\textbf{Generation of architectures}. For each target system, we executed each prompting strategy once, obtaining one architectural proposal in the ZS condition and one in the FS condition. No repeated executions were performed to estimate run-to-run variability. The API configuration and prompt templates were kept unchanged across the two systems within each condition, and the exact outputs were recorded in the replication package. The outputs were then manually normalized into the same representation used for the reference architectures, preserving the proposed services, responsibilities, and communication links while standardizing only their presentation. Consequently, the quantitative values reported in this study describe the observed executions rather than a distribution of possible model outputs.

This step produced three candidate architectures per system: the implemented reference architecture, the ZS LLM-generated architecture, and the FS LLM-generated architecture. These architectures are then used in the next step to perform a quantitative comparison against the reference systems, enabling the assessment of service identification and interaction recovery.

\subsubsection*{Step 3: Quantitative and Comparative Evaluation}
\label{subsec:quantitative_evaluation}

The third step performs a quantitative comparison between the LLM-generated architectures and the reference decompositions. Its objective is to assess how accurately the model recovers both the set of services and the communication relationships among them, providing a structured evaluation of architectural correspondence.

\textbf{Comparison of Identified Services}. The first stage evaluates service identification (RQ1) by comparing the set of services proposed by the model with those defined in the reference architecture. For each generated service, we assess whether a corresponding reference service exists with similar responsibilities and functional scope. This comparison considers service names and their semantic roles, including the business capabilities they encapsulate and the operations they are expected to support.

To perform this matching, we analyze service descriptions and responsibilities to determine whether a generated service captures the same architectural function as a reference service. This allows us to identify whether the model correctly decomposes the system into meaningful and coherent service boundaries.

\textbf{Comparison of Inter-Service Interactions}. The second stage evaluates interaction alignment (RQ2) by comparing the communication links among services. Each interaction proposed by the model—typically represented as a directed connection between two services—is matched against the set of interactions present in the reference architecture.

The comparison focuses on whether the model correctly identifies which services should communicate and the directionality of these interactions. As with service matching, the evaluation considers semantic equivalence rather than exact representation, ensuring that interactions are assessed based on their architectural intent (e.g., data exchange, coordination, or dependency) rather than strictly on naming or notation differences.

\textbf{Classification protocol}. For service identification, each service proposed by the model was compared against the list of services in the reference architecture. Services were classified into three categories:

\begin{itemize}
    \item \textit{Correct}: services that correspond to a reference service in scope and responsibility;
    \item \textit{Missing}: reference services not recovered by the model;
    \item \textit{Extra}: services proposed by the model that do not correspond to a reference service.
\end{itemize}

For interaction alignment, the same logic was applied to communication links. Each proposed interaction was compared against the set of communication relationships present in the reference architecture and classified as correct, missing, or extra.

Because architectural naming and granularity can vary across decompositions, the comparison was based on literal service names and the services' semantic roles and responsibilities. In cases where naming differed but responsibility alignment was clear, correspondence was determined based on architectural function rather than exact lexical identity. This rule was necessary to avoid unfairly penalizing outputs that correctly captured the business responsibility by using alternative labels. The matching procedure was performed by manually inspecting the generated and reference architectures against these criteria.

\textbf{Illustrative matching example}. Suppose that a generated architecture contains a \textit{Customer Management Service} responsible for customer registration and profile maintenance, whereas the reference contains a \textit{Customer Service} with the same responsibility. The generated service is classified as \textit{correct} despite the lexical difference because its architectural scope is semantically equivalent. If the model additionally proposes a \textit{Notification Service} that has no counterpart in the reference, it is classified as \textit{extra} relative to that baseline; if an implemented \textit{Payment Service} is not recovered, it is classified as \textit{missing}. For interactions, a generated directed link \textit{Order Service $\rightarrow$ Inventory Service} is correct only when the same semantically aligned dependency and direction are present in the reference. An unsupported or reversed-only link is extra, whereas an expected but absent link is missing. This example illustrates the protocol and does not correspond to a new empirical result.

\textbf{Metrics calculation}. For both services and interactions, we computed \textit{precision}, \textit{recall}, and \textit{F1-score}. Let $TP$ denote the number of correctly identified elements, $FP$ the number of extra elements, and $FN$ the number of missing elements. Then:

{
\small

\begin{equation}
Precision = \frac{TP}{TP + FP}
\end{equation}

\begin{equation}
Recall = \frac{TP}{TP + FN}
\end{equation}

\begin{equation}
F1 = 2 \cdot \frac{Precision \cdot Recall}{Precision + Recall}
\end{equation}

}

Precision indicates the proportion of generated elements that correspond to the implemented reference, while recall measures the proportion of reference elements recovered by the model. The F1-score summarizes this agreement when outputs may simultaneously omit elements and introduce alternatives. These metrics were computed separately for services and directed interactions. They operationalize RQ1 and RQ2 as structural correspondence with a particular implementation; they do not, by themselves, establish intrinsic architectural quality or prove that every non-reference element is incorrect. This distinction motivates the complementary expert evaluation.

At the end of this step, we obtain quantitative metrics that characterize both the accuracy and the types of errors present in the generated architectures. These results provide the basis for subsequent expert validation, in which the perceived quality and practical plausibility of the architectures are assessed.

\subsubsection*{Step 4: Expert Validation}
\label{subsec:expert_validation}

The final step aimed to evaluate the perceived quality, correctness, and practical viability of the generated architectures from the perspective of experienced software architects. While the previous step measures structural correspondence with the reference systems, expert validation addresses RQ3 by assessing whether the architectures are understandable, coherent, and credible in practice, and RQ4 by eliciting qualitative insights into recurring strengths, weaknesses, and design patterns observed in the generated solutions.

\textbf{Expert recruitment and evaluation materials}. We engaged six domain experts, each with at least five years of professional experience in designing distributed software systems. The inclusion criterion was intended to ensure that participants had sufficient practical familiarity with modular design, service separation, and architectural trade-offs to assess the alternatives meaningfully.

Each expert received an evaluation package for both target systems (i.e., PetClinic and Bookstore). For each system, the package included:

\begin{itemize}
    \item the full set of textual requirements;
    \item three architecture descriptions and diagrams; and
    \item a structured evaluation form.
\end{itemize}

The three architecture variants corresponded to: (i) the implemented reference architecture, (ii) the ZS LLM-generated architecture, and (iii) the FS LLM-generated architecture. To reduce bias, the three variants were anonymized and labeled generically as A, B, and C. Their order was randomized for each expert, so that participants could not infer which architecture was the implemented baseline or which one was generated by the model.

This blinding strategy was important because it prevented evaluators from favoring the reference architecture simply because it was known to be implemented, or from penalizing an architecture because it was known to be AI-generated.

The expert-based evaluation was designed to assess both the perceived architectural quality of the generated solutions and the qualitative insights derived from expert judgment. To this end, the evaluation was structured into two complementary components: a quantitative assessment and a qualitative analysis.

\textbf{Quantitative assessment}. Experts evaluated each architecture using a structured review form based on a 5-point Likert scale (1 = Very Poor, 5 = Excellent). Because these terms have multiple meanings in the literature, we adopted the following study-specific operational definitions:

\begin{itemize}
    \item \textit{Correctness}: the extent to which the proposed services, assigned responsibilities, and interactions are traceable to the stated requirements and do not contradict them. Unsupported responsibilities or dependencies reduce correctness;
    \item \textit{Completeness}: the extent to which the architecture represents the functional capabilities, business responsibilities, and necessary coordination implied by the complete requirement set. Omitted required capabilities reduce completeness;
    \item \textit{Modularity}: the extent to which services exhibit cohesive responsibilities, clear ownership, separation of concerns, and limited unnecessary coupling. Blurred boundaries, duplicated ownership, excessive dependencies, or avoidable cycles reduce modularity;
    \item \textit{Plausibility}: the extent to which the architecture constitutes an implementable and operationally credible microservice solution, considering realistic dependency directions, data and responsibility allocation, deployment independence, and the treatment of cross-cutting concerns.
\end{itemize}

Thus, correctness and completeness assess requirements alignment from different perspectives (validity of included decisions versus coverage of required capabilities), while modularity and plausibility assess structural quality and real-world feasibility. The scale anchors indicate the degree to which each architecture satisfies the corresponding operational definition.

The evaluation procedure followed a controlled and blinded protocol. Experts reviewed each system independently: they first analyzed the textual requirements and then evaluated the three architectural alternatives using the same form and criteria. The alternatives were presented under identical conditions, enabling direct comparison. The assessment was conducted individually to avoid discussion effects and preserve independent judgment.

The ratings were retained at the individual-expert level and consolidated descriptively by system, architecture variant, and evaluation dimension. We did not conduct a consensus meeting or replace divergent scores with an adjudicated value; disagreement was preserved as part of the evidence about architectural trade-offs. The resulting summaries support comparison among the alternatives for RQ3, but the six-expert sample is not used for population-level statistical generalization.

\textbf{Qualitative assessment}. In addition to quantitative ratings, experts provided open-ended feedback to capture deeper architectural reasoning. Specifically, they addressed the following aspects:

\begin{itemize}
    \item What are the main strengths of this architecture?
    \item What weaknesses or risks do you identify?
    \item Were there any surprising or non-obvious design choices?
    \item What modifications would you recommend to improve this architecture?
\end{itemize}

The open-ended responses were analyzed through a structured descriptive synthesis rather than a formal grounded-theory or full thematic-analysis protocol. First, responses were organized by target system, anonymized architecture variant, and questionnaire question. Second, semantically similar observations were grouped into recurring descriptive categories (e.g., excessive coupling, unclear responsibility allocation, unsupported infrastructure services, or coordination inefficiencies). Third, the categories were compared across the reference, ZS, and FS alternatives to identify repeated and contrasting assessments. Individual comments that disagreed with the dominant pattern were retained and considered in the interpretation rather than being forced into consensus. No inter-rater reliability statistic was calculated for this author-led synthesis. The procedure supports RQ4 by making the consolidation process transparent and helps interpret the RQ3 ratings, while its interpretive nature is acknowledged as a reliability limitation.

\section{Results}
\label{sec:results}

This section reports the empirical results according to the four research questions introduced in Section~\ref{sec:methodology}. We first present structural agreement for service identification (Section 5.1) and inter-service communication recovery (Section 5.2). We then report the expert ratings of perceived architectural quality (Section 5.3) and the structured descriptive synthesis of recurring strengths, weaknesses, and design patterns (Section 5.4). The experiments were conducted in May and June 2025 using OpenAI o3 via the OpenAI API. Each reported ZS or FS architecture corresponds to one execution, so the values characterize these recorded outputs and should not be interpreted as estimates of run-to-run stability. All artifacts---including prompts, generated outputs, normalized architectures, and evaluation materials---are available in the supplementary repository~\cite{dataset}.

\subsection{Identification of Microservice Components from Requirements (RQ1)}
\label{subsec:results_rq1}

\begin{figure}[!ht]
    \centering
    \includegraphics[width=0.99\linewidth]{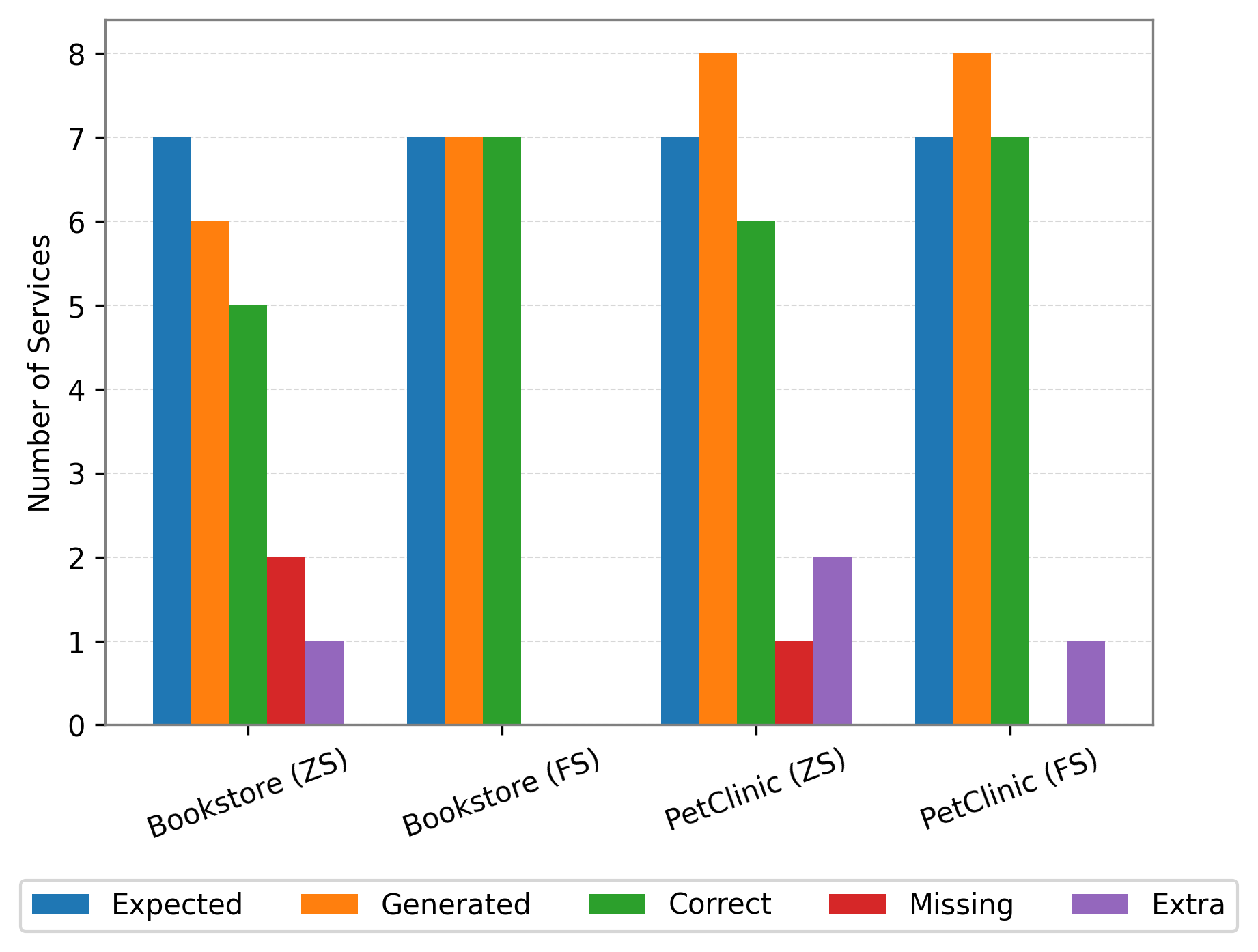}
    \caption{Evaluation of services identified by prompt strategies.}
    \label{fig:rq1_services}
\end{figure}

RQ1 investigates whether the LLM can infer the microservice components of a system directly from natural-language requirements. Figure~\ref{fig:rq1_services} summarizes the results for both case studies and prompting conditions.

\textbf{Analysis of raw results}. The raw counts reveal distinct behaviors across prompting strategies. In the \textit{Bookstore system}, the ZS condition generated six services, correctly identifying five of the seven expected services, missing two, and introducing one additional service. In contrast, the FS condition produced exactly the seven expected services, with no omissions or extra components, achieving perfect alignment with the reference decomposition.

In the \textit{PetClinic system}, which exhibits higher domain complexity, the ZS condition generated eight services, correctly identifying six of the seven expected services, missing one, and introducing two additional services. Under FS prompting, the model again generated eight services, successfully recovering all seven expected services while introducing only one additional service. 

Across both systems, ZS produced 11 correct services out of 14, with three missing and three extra services overall. In contrast, FS recovered all 14 expected services and introduced only one additional service. These results indicate that FS consistently improves coverage while reducing structural inconsistencies in the generated architectures.

\begin{figure}[!ht]
    \centering
    \includegraphics[width=0.95\linewidth]{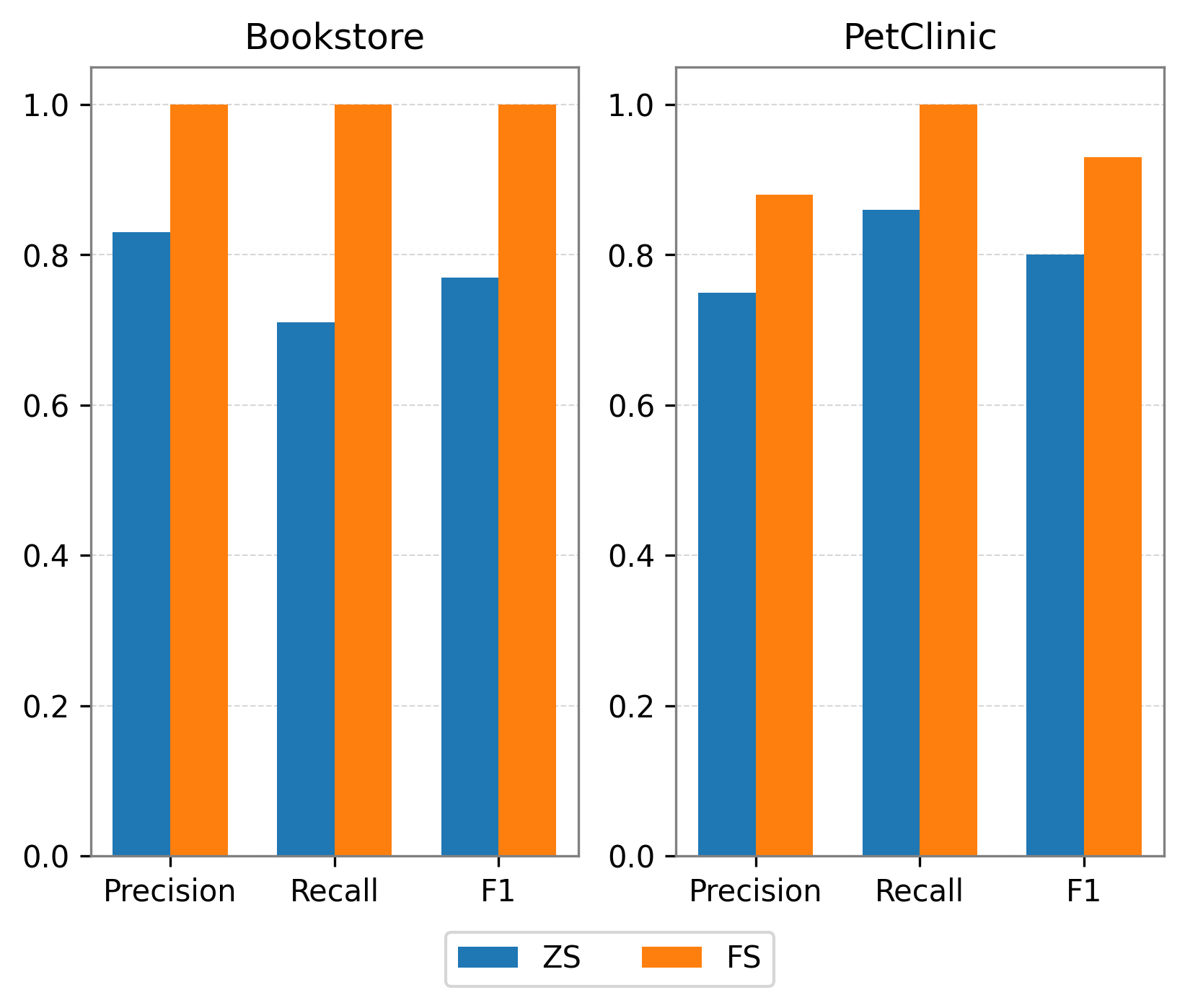}
    \caption{Precision, Recall, and F1-score for service identification across prompting strategies.}
    \label{fig:rq1_metrics}
\end{figure}

\textbf{Metric-based evaluation}. Figure~\ref{fig:rq1_metrics} presents the corresponding precision, recall, and F1-score values.

The metric analysis confirms the trends observed in the raw data. 
In the \textit{Bookstore system}, the F1-score increases from 0.77 under ZS to 1.00 under FS, reflecting both complete recovery and the absence of spurious services. In the \textit{PetClinic system}, the F1-score increases from 0.80 to 0.93, indicating improved completeness and a reduction in over-segmentation.

At the aggregated level, ZS achieves a precision of 0.79, a recall of 0.79, and an F1-score of 0.79. In contrast, FS achieves a precision of 0.93, perfect recall of 1.00, and an F1-score of 0.97. These improvements demonstrate that FS enhances both dimensions simultaneously, increasing coverage while constraining unnecessary service generation.

\textbf{Joint analysis and interpretation}. Taken together, the raw and metric-based results reveal a consistent pattern. Under ZS, the model exhibits two types of errors: (i) missing services, indicating incomplete coverage of the domain, and (ii) extra services, suggesting over-segmentation of responsibilities. These behaviors vary with system complexity but consistently yield moderate precision and recall.

In contrast, FS significantly reduces both types of errors. The elimination of missing services across both systems indicates that the model becomes more sensitive to requirement coverage, while the reduction of extra services suggests improved control over granularity. This indicates that the FS example provides implicit structural guidance, helping the model calibrate the decomposition process.

\medskip
\noindent\fcolorbox{gray!80}{white}{%
\parbox{0.98\linewidth}{%
\textbf{Answer to RQ1.} 
The results indicate that LLMs can infer microservice components directly from natural-language requirements with high effectiveness. While ZS prompting yields moderately accurate decompositions, FS prompting substantially improves both completeness and precision, enabling near-perfect recovery of service boundaries across different systems. This suggests that minimal exemplar-based guidance is sufficient to transform LLM outputs into reliable architectural decompositions.
}}

\subsection{Recovery of Inter-Service Communication Links (RQ2)}
\label{subsec:results_rq2}

\begin{figure}[!ht]
    \centering
    \includegraphics[width=0.99\linewidth]{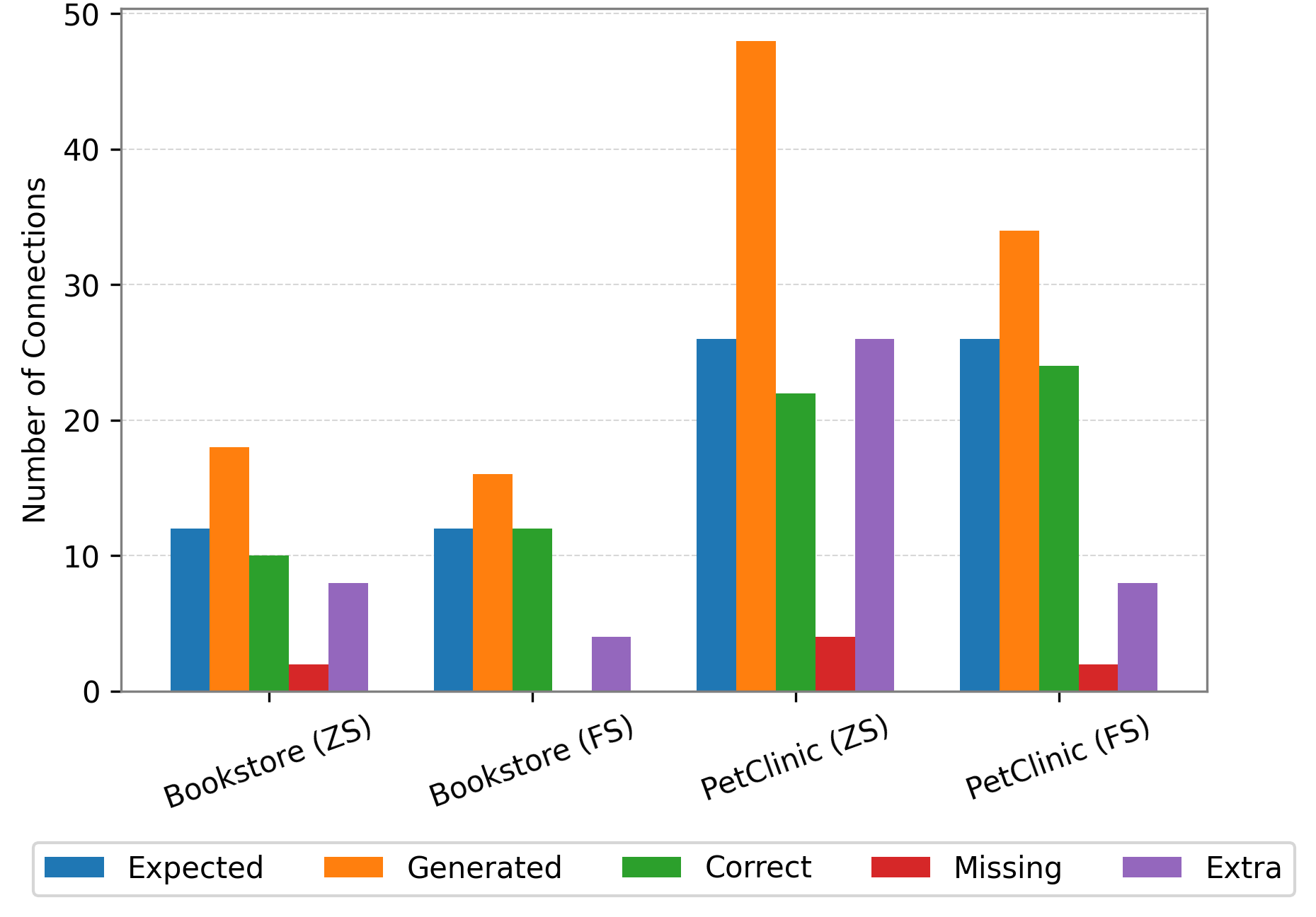}
    \caption{Inter-Service communication identified by prompt strategies.}
    \label{fig:rq2_interactions}
\end{figure}

RQ2 evaluates whether the LLM can recover the communication relationships among services defined in the reference architectures. Figure~\ref{fig:rq2_interactions} reports the interaction-level results.

\textbf{Analysis of raw results}. The raw interaction counts reveal a consistent pattern across both systems, characterized by high coverage but substantial over-generation under the ZS condition. 

In the \textit{Bookstore system}, ZS produced 18 interactions, correctly matching 10 of the 12 expected links, missing two, and introducing eight additional connections. Under FS prompting, all 12 expected interactions were recovered, and the number of extra links was reduced to four. 

In the \textit{PetClinic system}, which involves a more complex interaction structure, ZS generated 48 interactions, correctly identifying 22 of the 26 expected links, missing four, and introducing 26 additional connections. Under FS prompting, the model generated 34 interactions, correctly matching 24 of the 26 expected links, with two missing and 10 additional links.

Across both systems, ZS correctly recovered 32 of the 38 reference interactions, with 6 missing and 34 extra links overall. In contrast, FS recovered 36 of the 38 interactions, reducing missing links to 2 and extra links to 14. These results indicate that while both strategies achieve high coverage, FS significantly reduces the number of spurious interactions.

\begin{figure}[!ht]
    \centering
    \includegraphics[width=0.95\linewidth]{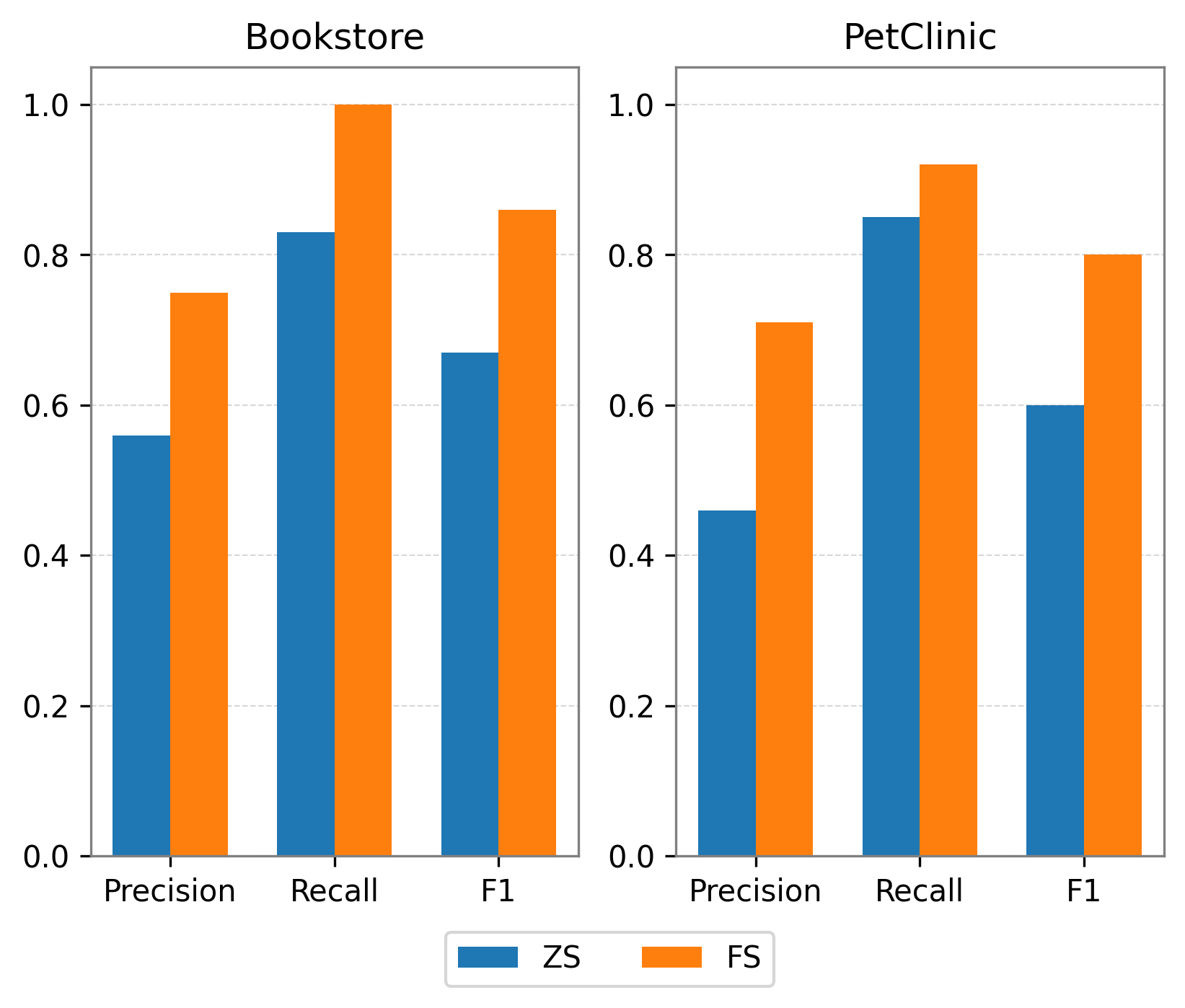}
    \caption{Precision, Recall, and F1-score for inter-services communication across prompting strategies.}
    \label{fig:rq2_metrics}
\end{figure}

\textbf{Metric-based evaluation}. Figure~\ref{fig:rq2_metrics} presents the corresponding precision, recall, and F1-score values.

The metric analysis confirms the trends observed in the raw data. In the \textit{Bookstore system}, the F1-score increases from 0.67 under ZS to 0.86 under FS, reflecting both improved completeness and reduced over-generation. In the \textit{PetClinic system}, the F1-score increases from 0.59 to 0.80, indicating that FS improves alignment even in more complex interaction graphs, where the number of potential dependencies is higher and more difficult to infer accurately.

At the aggregated level, ZS achieves a precision of 0.48, a recall of 0.84, and an F1-score of 0.61. In contrast, FS achieves a precision of 0.72, a recall of 0.95, and an F1-score of 0.82. These results highlight a key characteristic of the interaction-recovery task: recall remains consistently high across both strategies, suggesting that the model generally captures relevant interactions, while precision varies substantially with the level of over-generated connections. This indicates that the primary challenge lies not in identifying potential interactions but in filtering out those that do not meet the requirements.

\textbf{Joint analysis and interpretation}. Taken together, the results reveal a systematic behavior of LLMs in interaction inference. Under ZS prompting, the model tends to produce densely connected architectures, recovering most expected interactions but introducing many unsupported links. This results in high recall but low precision, particularly evident in the PetClinic system.

In contrast, FS prompting significantly reduces over-connectivity while preserving high recall. Although some extra links remain, especially in the more complex system, the reduction is substantial, leading to a more balanced trade-off between completeness and correctness. This suggests that FS examples act as implicit constraints on interaction patterns, helping the model avoid excessive connectivity and produce more realistic architectural structures.

Importantly, the interaction-recovery task remains more challenging than service identification. Even under FS prompting, the model continues to introduce additional links, indicating that inferring precise interaction structures from textual requirements requires not only identifying relevant services but also reasoning about coordination patterns, data flow, and dependency necessity. 

\medskip
\noindent\fcolorbox{gray!80}{white}{%
\parbox{0.98\linewidth}{%
\textbf{Answer to RQ2.} 
The results indicate that LLMs can recover most inter-service interactions from textual requirements, achieving consistently high recall across both prompting strategies. However, ZS prompting tends to yield overly dense interaction graphs with many unsupported links, leading to low precision. FS prompting substantially improves this behavior by reducing over-generated connections while maintaining high coverage, leading to more balanced and plausible interaction structures. Nevertheless, interaction recovery remains more challenging than service identification, particularly in systems with higher structural complexity.
}}

\subsection{Quantitative Assessment of Architectural Quality by Experts (RQ3)}
\label{subsec:results_rq3}

RQ3 examines how software architecture experts evaluate the generated architectures based on structured quantitative criteria. The evaluation was conducted in a blinded setting, where three architectural alternatives were presented anonymously as Architectures A (reference), B (ZS), and C (FS). Experts assessed each alternative across four dimensions: correctness, completeness, modularity, and plausibility, using a five-point Likert scale.

\begin{figure*}[!ht]
\centering
\includegraphics[width=0.90\linewidth]{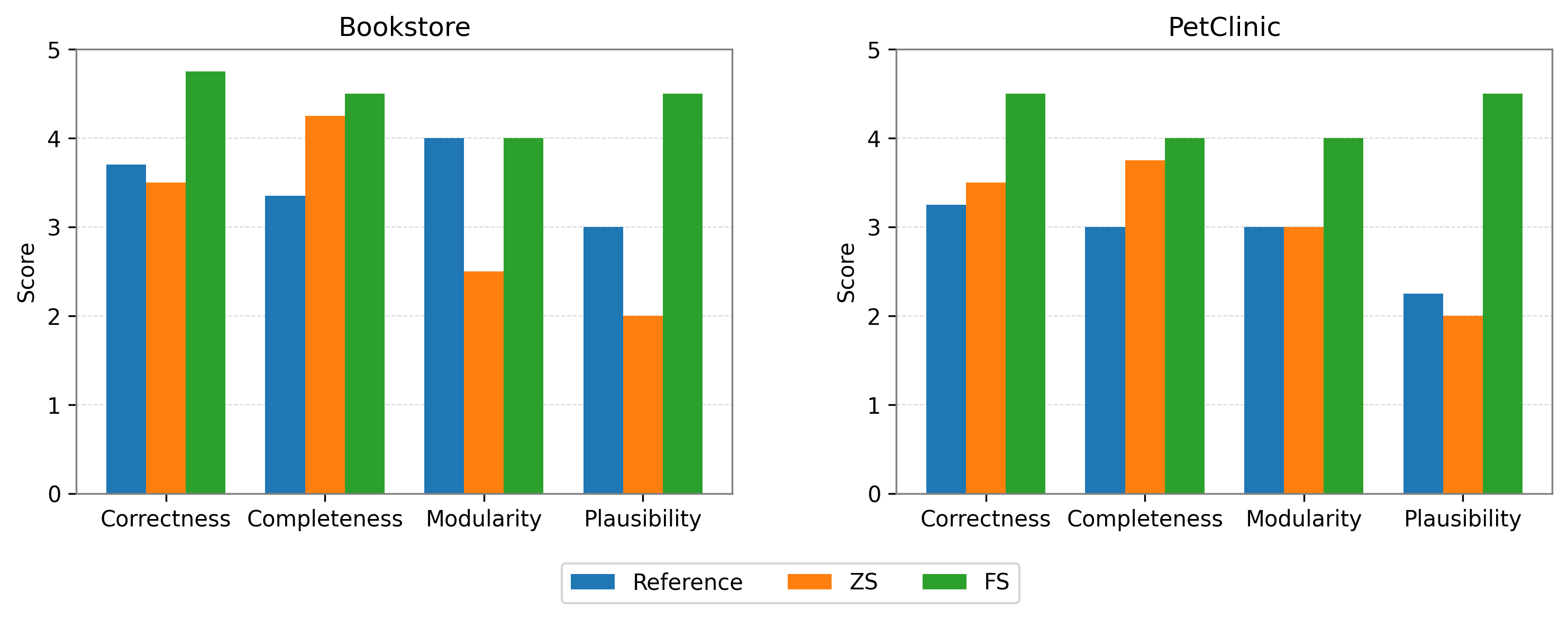}
\caption{Average expert ratings for RQ3.}
\label{fig:rq3_expert_scores}
\end{figure*}

Figure~\ref{fig:rq3_expert_scores} summarizes the average scores assigned by the experts for both systems.

Across both systems, the FS architectures consistently achieved the highest scores in nearly all evaluation dimensions. In the Bookstore system, FS reached 4.75 in correctness, 4.50 in completeness, 4.00 in modularity, and 4.50 in plausibility. A similar pattern was observed in the PetClinic system, where FS achieved 4.50, 4.00, 4.00, and 4.50, respectively. These results indicate that FS improves structural alignment with the requirements (correctness and completeness), and enhances perceived design quality (modularity and plausibility).

The superior performance of FS architectures can be interpreted as a direct consequence of the structural guidance provided by exemplar-based prompting. Unlike ZS, which relies solely on implicit reasoning, FS appears to constrain the architectural search space, leading to solutions that are both comprehensive and structurally coherent. This aligns with the improvements observed in RQ1 and RQ2, where FS achieved near-perfect service identification and significantly reduced spurious interactions. Together, these findings suggest that FS enables the model to better balance coverage and organization.

On the other hand, the ZS architectures exhibited a distinct quantitative profile characterized by high completeness but low plausibility and modularity. In both systems, ZS achieved relatively high completeness scores (4.25 in Bookstore and 3.75 in PetClinic), often exceeding the reference architectures. This indicates that ZS is effective at capturing a broad set of functional elements from the requirements. However, this gain is accompanied by a significant degradation in structural quality. Plausibility scores dropped to 2.00 in both systems, and modularity was notably lower (e.g., 2.50 in Bookstore). This divergence reflects a fundamental trade-off: ZS tends to prioritize coverage at the expense of architectural discipline, leading to overly dense, highly coupled designs. These results are consistent with RQ2, which found that ZS exhibited high recall but low precision due to excessive interaction generation.

The reference architectures occupied an intermediate position across all dimensions. They achieved relatively strong modularity (e.g., 4.00 in Bookstore), indicating well-structured and coherent designs, but lower completeness compared to ZS and FS. This suggests that the implemented architectures are conservative in scope, prioritizing stability and clarity over exhaustive requirement coverage. Interestingly, this pattern highlights a key contrast between human-designed and LLM-generated architectures: while human-designed systems tend to avoid unnecessary complexity, LLM-generated solutions—especially under ZS—tend to over-generalize, introducing additional elements and interactions.

The observed patterns remain consistent across both case studies, despite differences in domain complexity. In both Bookstore and PetClinic, FS dominates across all dimensions, ZS shows a clear imbalance between completeness and structural quality, and the reference architectures remain stable but less comprehensive. This consistency strengthens the validity of the findings and suggests that the effects of prompting strategies are robust across different domains.

Importantly, these results provide insight into how LLMs perform architectural reasoning under different prompting conditions. While ZS prompting encourages aggressive requirement-to-structure mapping and prioritizes coverage, FS prompting introduces implicit structural constraints that guide the model toward more coherent and modular designs. This highlights the critical role of contextual grounding in enabling effective architectural reasoning with LLMs.

\medskip
\noindent\fcolorbox{gray!60}{white}{%
\parbox{0.98\linewidth}{%
\textbf{Answer to RQ3.}
The quantitative expert evaluation indicates that FS prompting produces the highest-quality architectural proposals across correctness, completeness, modularity, and plausibility. While ZS prompting tends to increase completeness, it does so at the expense of plausibility and modularity. The reference architectures exhibit balanced but less comprehensive designs. Overall, FS prompting yields the most consistently well-rated architectures across both systems.
}}

\subsection{Qualitative Assessment of Architectural Strengths and Weaknesses (RQ4)}

RQ4 investigates how experts perceive the architectural trade-offs introduced by different prompting strategies, focusing on recurring strengths, weaknesses, design patterns, and improvement suggestions. While RQ1 and RQ2 quantify structural alignment and RQ3 captures perceived quality through ratings, this analysis provides a deeper interpretation of \textit{why} certain architectures are preferred, revealing how prompting influences architectural reasoning.

To support this analysis, we examine expert feedback across two systems (Bookstore and PetClinic) and organize the discussion around three architectural variants (i.e., Reference, ZS, and FS). For each case, we analyze qualitative responses across strengths, weaknesses, surprising design choices, and suggested improvements.

\begin{figure*}[!ht]
\centering
\includegraphics[width=0.85\linewidth]{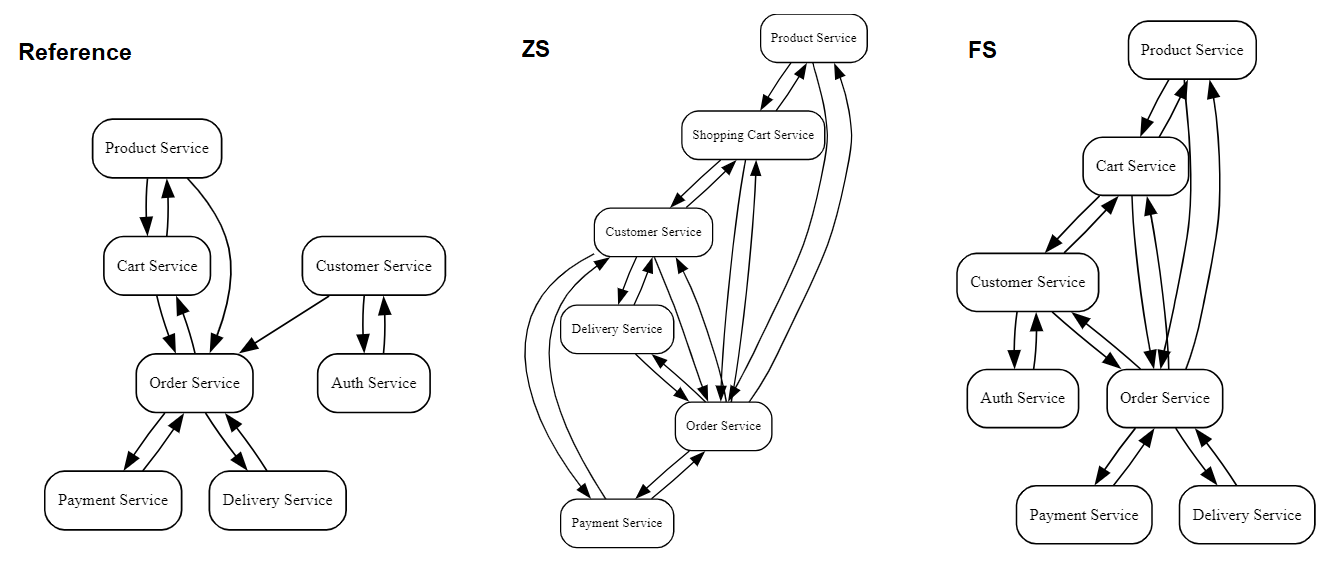}
\caption{Comparison between the architectural variants for the Bookstore system.}
\label{fig:arch_bookstore}
\end{figure*}

\textbf{Bookstore system}. Figure~\ref{fig:arch_bookstore} provides a structural overview of the three architectural alternatives for the Bookstore system, highlighting clear differences in both the number of services (RQ1) and the density of inter-service connections (RQ2).

At a high level, the reference architecture exhibits a compact and centralized structure, with a limited number of services and a relatively sparse interaction graph. This aligns with the results of RQ1 and RQ2, which showed that the reference solution achieved balanced service identification and a controlled number of interactions, resulting in a structurally coherent but functionally restrained design.

In contrast, the ZS architecture presents a visibly denser and more complex interaction topology. Although it introduces additional services—consistent with the slight over-generation observed in RQ1—the most prominent difference lies in the number and directionality of connections. The interaction graph is highly interconnected, with multiple bidirectional links and cross-service dependencies, reflecting the high number of extra interactions identified in RQ2. This results in an architecture that is functionally expressive but structurally overloaded.

The FS architecture occupies an intermediate position between these two extremes. While maintaining a service set comparable to the reference (RQ1), it introduces a richer interaction structure without reaching the level of over-connectivity observed in ZS (RQ2). The resulting topology appears more organized, with clearer communication paths and fewer redundant links, suggesting a more controlled expansion of the architecture.

Taken together, these visual differences reinforce the quantitative findings from RQ1 and RQ2: while ZS tends to increase both service and interaction counts, leading to dense, potentially overconnected architectures, FS achieves a more balanced structure, preserving coverage while improving interaction organization.

\textbf{Qualitative analysis of expert feedback}. The qualitative responses were analyzed across four dimensions: strengths, weaknesses, surprising design choices, and suggested improvements. These findings complement the quantitative results of RQ3 by explaining the patterns observed in expert ratings.

\underline{Reference architecture}. The reference architecture was consistently described as simple, well-structured, and easy to understand. Experts highlighted the clear separation of responsibilities and the Order Service's central role in coordinating the purchasing workflow. Low coupling and straightforward interactions were perceived as key strengths, as illustrated by comments such as \textit{``simple modules, low coupling''} and \textit{``clear and understandable service definitions''}. 

These perceptions are consistent with the relatively strong modularity scores observed in RQ3. However, several limitations were identified, particularly regarding functional coverage and service integration. Reviewers noted that some services were weakly connected or absent from the main workflow (e.g., \textit{``Auth Service does not participate in the cart or order flow''}), which helps explain the lower completeness scores assigned to the reference architecture. The concentration of logic in the Order Service (e.g., \textit{``too much logic in Order Service''}) was also identified as a potential structural risk.

\underline{ZS architecture}. This architecture was consistently associated with strong functional coverage, which aligns with its higher completeness scores in RQ3. Experts noted that most relevant services were present and connected (e.g., \textit{``complete coverage of functional requirements''}), indicating that the model captured a broad set of domain responsibilities.

However, this increased coverage came at the cost of structural quality. Reviewers frequently described excessive coupling and unnecessary interactions, with comments such as \textit{``Most microservices are coupled with each other''} and \textit{``relationships between services seem unnecessary''}. These observations provide a direct explanation for the lower modularity and plausibility scores observed in RQ3. Additional concerns regarding maintainability and scalability (e.g., \textit{``difficult to scale'' and ``It will be hard to maintain''}) further reinforce the interpretation that ZS produces architectures that are functionally complete but structurally overloaded.

\underline{FS architecture}. This architecture received the most balanced qualitative evaluation, which is consistent with its superior performance across all quantitative dimensions in RQ3. Experts highlighted clearer service boundaries and more coherent interaction flows (e.g., \textit{``clear functional boundaries and clearer flows''}), indicating improved structural organization compared to ZS.

At the same time, FS maintained broad functional coverage, contributing to its high completeness scores. This combination of coverage and structural clarity explains the consistently high ratings across correctness, modularity, and plausibility. Nevertheless, some residual issues were still identified, particularly related to remaining dependencies in central services (e.g., \textit{``Order remains over-connected''}) and unexpected design elements (e.g., \textit{``Auth service appears without an explicit requirement''}). These observations help contextualize why FS, while superior, does not fully eliminate all structural concerns.

\begin{figure*}[!ht]
\centering
\includegraphics[width=0.95\linewidth]{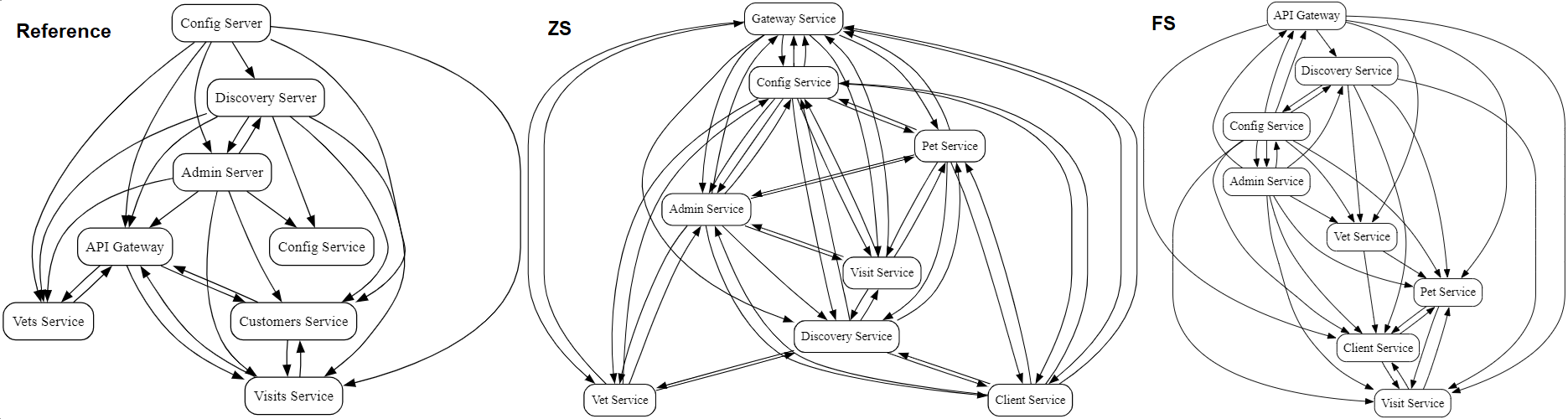}
\caption{Comparison between the architectural variants for the PetClinic system.}
\label{fig:arch_petclinic}
\end{figure*}

\textbf{PetClinic system}. Figure~\ref{fig:arch_petclinic} presents the architectural variants for the PetClinic system. Compared to the Bookstore case, this system exhibits a richer domain and a more complex interaction topology, making it a more demanding scenario for architectural inference.

In the PetClinic case, the reference architecture was described as simple and restrained, with a clear overall structure. However, several reviewers questioned missing service responsibilities and limited functional decomposition, suggesting that the architecture underrepresented some domain aspects.

The ZS architecture was again associated with both high functional coverage and structural complexity. Experts frequently reported excessive coupling, gateway-centered communication patterns, and bidirectional flows, all of which increased architectural complexity. The presence of auxiliary or loosely defined services was also noted, raising concerns about clarity and maintainability.

The FS architecture was evaluated as the most balanced alternative. Reviewers highlighted improved organization of veterinary-domain services, clearer separation between responsibilities, and more consistent interaction patterns. Although some additional links remained, the architecture was generally perceived as more coherent and operationally plausible than the ZS variant.

\textbf{Qualitative analysis of expert feedback}. The qualitative responses were analyzed across four dimensions: strengths, weaknesses, surprising design choices, and suggested improvements. These findings also complement the quantitative results of RQ3 by explaining the patterns observed in expert ratings.

\underline{Reference architecture}. The reference architecture was generally perceived as simple and relatively easy to understand. Experts highlighted the presence of \textit{``Services are well identified''} and a \textit{``The call flow seems realistic and consistent''}, suggesting that the architecture provides a coherent and interpretable interaction structure. The relatively low number of connections was also considered a strength, as reflected in comments such as \textit{``low number of unnecessary connections''} and \textit{``few modules (microservices)''}, indicating a lean and straightforward design.

These characteristics are consistent with the moderate-to-high modularity scores observed in RQ3. However, several limitations were identified. A recurring concern was the lack of important domain services, particularly the absence of a Pet Service (e.g., \textit{``No pet service created in the diagram''} and \textit{``Services like Pet Service are lacking''}). Experts also noted excessive reliance on infrastructural components, especially the Admin Server (e.g., \textit{``Heavy reliance on Admin Server''} and \textit{``Configuration Service is only available through the admin server''}). Additionally, some responsibilities appeared duplicated or unclear (e.g., \textit{``Modules with apparently duplicate responsibilities''}), and certain interactions were considered poorly defined. These issues help explain the lower completeness scores observed in RQ3, as the architecture, while simple, fails to fully capture the domain structure.

\underline{ZS architecture}. This architecture was consistently associated with strong functional coverage. Experts emphasized that \textit{``All required services are available''} and \textit{``all required services identified''}, indicating that the model successfully captured most domain elements. The placement of the API Gateway was also considered appropriate (e.g., \textit{``API Gateway in correct position''}), reinforcing the perception of improved structural alignment compared to the reference.

However, this improved coverage came at the cost of structural quality. A dominant issue across evaluations was excessive coupling, with multiple experts highlighting that \textit{``all the microservices are unnecessarily connected''} and \textit{``excessive coupling''}. The architecture was frequently described as overly dense, with comments such as \textit{``connecting everything with everything''} and \textit{``All the services seem to be calling each other''}.

Another critical issue concerns the misuse of architectural roles. Experts noted that the Gateway was being invoked internally (e.g., \textit{``Gateway receives calls from other services''} and \textit{``Gateway invoked internally, which breaks its entry-point role''}), which violates its expected behavior. Similarly, Admin and Config services were seen as excessively central (e.g., \textit{``They appear in the middle of everything, which is neither necessary nor scalable''}). These issues directly explain the lower modularity and plausibility scores observed in RQ3, indicating that while ZS improves completeness, it produces architectures that are difficult to maintain, scale, and justify structurally.

\underline{FS architecture}. This architecture received the most balanced qualitative evaluation, aligning with its superior performance in RQ3. Experts highlighted improvements in structural clarity and organization, with comments such as \textit{``cleaner and more controlled flow''}, \textit{``cleaner, scoped flow with a single-entry gateway''}, and \textit{``Good separation between Client, Pet, Visit, and Vet''}. The correct positioning of the API Gateway as a single entry point was also emphasized (e.g., \textit{“API Gateway correctly positioned as main entry”}), reinforcing a more disciplined architectural structure.

In addition to improved modularity, FS maintained adequate functional coverage, which explains its high completeness and correctness scores. The architecture also showed reduced coupling compared to ZS (e.g., \textit{``relatively low coupling among the identified microservices''}), indicating better control over service interactions.

Nevertheless, some residual issues were still identified. Experts noted remaining unnecessary dependencies, particularly involving the Admin Service (e.g., \textit{``Admin still over-connected''} and \textit{``The Admin Service continues to have more connections than necessary''}). There were also concerns about unclear or unjustified links (e.g., \textit{``gateway $\leftrightarrow$ Config path unclear''} and \textit{``double arrow between Config and Admin feels out of scope''}). Additionally, some design elements, such as the inclusion of certain infrastructural services, were perceived as not directly derived from the requirements (e.g., \textit{``Inclusion of API Gateway, which doesn't seem to come strictly from the requirements''}).

These observations indicate that, although FS significantly improves architectural balance, it does not fully eliminate all inconsistencies. Still, compared to the other alternatives, it produces the most coherent and plausible architectural design.

\medskip
\noindent\fcolorbox{gray!60}{white}{%
\parbox{0.98\linewidth}{%
\textbf{Answer to RQ4.}
Qualitative feedback indicates that ZS prompting tends to produce architectures with high functional coverage but excessive coupling and limited plausibility. FS prompting, in contrast, leads to more modular, coherent, and interpretable architectures, while still maintaining broad coverage. These findings suggest that prompting strategies directly influence the structural quality of LLM-generated designs, with FS prompting yielding more balanced and realistic architectural solutions across different systems.
}}

\section{Analysis}
\label{sec:analysis}

This section deepens the interpretation of the original findings through a secondary analysis of the quantitative results, generated architectural artifacts, and expert assessments. Rather than introducing new experiments, it reexamines the evidence already reported in the study to better characterize prompting effects, architectural error patterns, and expert evaluation dynamics.

\textbf{Comparative Secondary Analysis of Prompting Effects}. A secondary reading of the quantitative results suggests that a key effect of FS prompting is not limited to higher accuracy. More fundamentally, it appears to introduce a form of structural discipline on the mapping between textual requirements and architectural decomposition. Across the two case studies, FS prompting produces a more consistent alignment with the reference architectures than ZS prompting, both in service identification and in inter-service communication recovery. In service identification, the reported F1 values increase from 0.77 to 1.00 in Bookstore and from 0.80 to 0.93 in PetClinic. In communication recovery, the same tendency is observed, with F1 increasing from 0.67 to 0.86 in Bookstore and from 0.59 to 0.80 in PetClinic.

This difference is analytically relevant because the gain is not uniform across tasks. The improvement is greater and cleaner for service discovery than for communication recovery. This indicates that the model is more reliable at inferring candidate service boundaries from requirements than at reconstructing the full service dependency topology. In other words, the question of what services should exist appears easier for the model than the question of how those services should interact. From an architectural standpoint, this distinction is important because it places LLMs in a stronger position for early decomposition support than for precise interaction design.

A cross-case comparison reinforces this interpretation. FS prompting improves both systems, but the improvement is more limited in the more interaction-heavy case. PetClinic still exhibits residual noise even under FS prompting, especially in communication-level recovery, suggesting that increasing architectural complexity amplifies the model’s tendency to introduce plausible but weakly justified dependencies. Thus, FS prompting should be interpreted less as a guarantee of correctness and more as a constraint mechanism that narrows the architectural search space.

\textbf{Taxonomy of Architectural Error Patterns}
The generated outputs also enable the identification of recurring categories of architectural error. A first pattern is the omission of low-salience responsibilities. Services that are weakly emphasized or only indirectly implied in the requirements are more likely to be omitted, particularly in ZS prompting. This suggests that the model privileges lexical prominence and central business entities when inferring service boundaries.

A second pattern is the introduction of plausible but unsupported services. The study reports cases in which the model generates components, such as administrative, auditing, configuration, notification, or infrastructure-related services, that are architecturally reasonable in general but not sufficiently grounded in the requirements. These are not arbitrary hallucinations; they are better understood as overextensions of learned microservice design priors.

A third pattern is topological inflation. Especially in ZS prompting, the model tends to create communication structures with too many edges, sometimes including bidirectional or cyclic dependencies. This is evident in the denser designs generated and is also reflected in expert criticism regarding maintainability, scalability, and role ambiguity. Such outputs may look comprehensive, but they weaken the separation of concerns and make the resulting architecture harder to justify operationally.
A fourth pattern is boundary blurring between domain services and platform concerns. Some generated architectures treat gateway or infrastructure-oriented components as if they were central business services, or mix cross-cutting concerns directly into core domain flows. This reduces conceptual clarity and affects modularity. Taken together, these patterns show that the model’s mistakes are structured rather than random. They emerge when requirements are underspecified, when generic architectural priors become dominant, and when prompting does not adequately constrain the space of acceptable decompositions.

\textbf{Cross-Case Analysis of Expert Perceptions}
The expert evaluation becomes more informative when interpreted as evidence of a quality trade-off, rather than merely a ranking. In both case studies, the FS architecture is the only generated variant that more consistently combines functional coverage with acceptable structural organization. By contrast, ZS outputs are not penalized primarily for lack of functionality, but for how they achieve coverage: excessive coupling, unclear service responsibilities, and implausible communication structures.

This point is especially important because the experts evaluated the architectures along four distinct dimensions: correctness, completeness, modularity, and plausibility. The pattern reported in the study suggests that higher completeness alone is not enough to make a generated architecture acceptable. Experts appear to value completeness only when it is delivered with coherent modular boundaries and realistic interaction design. This explains why architectures that seem richer in functionality may still receive lower plausibility assessments.

The reference architectures also play an analytically useful role. They are not necessarily idealized solutions, but real, implemented decompositions that reflect practical trade-offs. Their comparison with the generated variants shows that architectural acceptance is shaped by whether the expected functions are present and whether the decomposition remains understandable, maintainable, and operationally credible. The FS condition stands out precisely because it approximates that balance more closely than the ZS condition.

This distinction also affects the interpretation of precision, recall, and F1-score. A high score indicates close agreement with the selected implementation, not proof that the generated design is universally superior. Conversely, an element counted as extra may represent a plausible alternative boundary, infrastructure component, or interaction that is absent from the baseline. Such alternatives reduce structural precision by definition, but their architectural merit must be judged using requirement traceability and expert assessment. The quantitative and expert results are therefore complementary rather than interchangeable.

\section{Implications}
\label{sec:implications}

The findings of this study have implications for research, practice, and tool support at the intersection of software architecture, requirements engineering, and intelligent design support.

\textbf{Implications for research}. The findings provide insights into how LLMs perform in requirements-driven architectural design and highlight directions for improving their effectiveness. In particular, they reveal task-specific limitations, structured error patterns, and the role of prompting in shaping architectural outcomes.

\begin{itemize}
    \item \textit{Service identification vs. interaction modeling.} 
    The results suggest that identifying service boundaries is a more tractable task for LLMs than recovering inter-service communication structures. This indicates that decomposition and interaction modeling should be treated as distinct subproblems, potentially requiring different prompting strategies or additional constraints tailored to each task.

    \item \textit{Prompting as a control mechanism.} 
    FS prompting does more than improve accuracy; it appears to regularize architectural generation by constraining the search space, reducing omissions, and limiting unsupported additions. This highlights prompting as a mechanism that shapes architectural reasoning, rather than merely formatting model input.

    \item \textit{Structured error patterns.} 
    The observed errors—such as the omission of low-salience responsibilities, the introduction of plausible but unsupported services, and the over-generation of interactions—suggest that model outputs are influenced by learned architectural priors. This motivates future research on error-aware prompting strategies, richer exemplars, and domain-constrained generation approaches.

    \item \textit{Hybrid evaluation is necessary.} 
    Structural similarity metrics alone do not fully capture architectural quality. The results reinforce the need for hybrid evaluation approaches that combine quantitative alignment with expert judgment, enabling assessment of modularity, plausibility, and operational feasibility.
\end{itemize}

\textbf{Implications for practice}. The results indicate how practitioners can effectively use LLMs to support architectural design activities. They highlight both the potential of LLMs as assistants in early-stage decomposition and the conditions under which their outputs are more reliable and actionable.
\begin{itemize}
    \item \textit{Support for early-stage design.} 
    LLMs can assist practitioners in generating initial microservice decompositions directly from textual requirements, which is particularly valuable in early design stages where architectural artifacts are still limited or evolving.

    \item \textit{Few-shot prompting improves reliability.} 
    The results show that even a single well-structured exemplar can substantially improve architectural quality, increasing coverage while reducing overgeneration. This makes FS prompting a practical and low-cost strategy for improving LLM outputs in real-world settings.

    \item \textit{Human-in-the-loop is essential.} 
    Generated architectures should be treated as decision-support artifacts rather than final designs. While LLMs can accelerate decomposition, expert validation remains necessary to refine interaction structures, resolve ambiguities, and ensure architectural plausibility.

    \item \textit{Dependence on requirement quality.} 
    The effectiveness of LLM-generated architectures depends strongly on the clarity and consistency of input requirements. This reinforces the importance of well-structured requirement descriptions, clear domain terminology, and reduced ambiguity in requirement engineering practices.
\end{itemize}

\textbf{Implications for tool support}. The study identifies opportunities to incorporate LLM-based architectural synthesis into software design tools. These insights point to new ways of supporting interactive, requirements-driven architectural exploration within development environments.
\begin{itemize}
    \item \textit{Integration into design environments.} 
    LLM-based architectural synthesis can be integrated into modeling tools and development environments to automatically generate initial service decompositions from requirement documents, supporting faster design initialization.

    \item \textit{Interactive architectural exploration.} 
    Such tools can enable iterative refinement of architectures through natural-language interaction, allowing practitioners to explore alternative decompositions and adjust design decisions dynamically.

    \item \textit{Improved traceability.} 
    Integrating LLMs into design workflows may strengthen traceability between requirements and architectural artifacts, helping maintain alignment between business intent and system structure throughout the development lifecycle.
\end{itemize}

Overall, the findings position OpenAI o3 as a promising assistant for requirements-driven architectural exploration in the two bounded cases studied. Its primary observed value lies in accelerating early decomposition and expanding the design space, while its main limitation remains the need for human validation, particularly for interaction design and architectural plausibility. Generalization to other LLMs, larger systems, and more complex organizational settings requires independent replication.

\section{Threats to Validity}
\label{sec:threats}

We discuss potential threats to validity following the four dimensions commonly adopted in empirical software engineering: construct validity, internal validity, external validity, and reliability~\cite{wohlin2012experimentation}.

\textbf{Construct Validity}. 

A first construct-related threat concerns the use of the implemented architectures as reference baselines. The selected repositories provide concrete decompositions and dependency records, but they are not guaranteed to be optimal, exhaustive, or representative of industrial microservice practice. Static extraction can also miss dynamic or externally configured dependencies. We therefore treat precision, recall, and F1-score as measures of agreement with these particular implementations rather than measures of architectural quality. Blinded expert evaluation was added to assess whether generated alternatives---including elements counted as extra relative to the baseline---were perceived as correct, complete, modular, and plausible.

A second construct-related threat concerns how correspondence was established between the generated and reference architectures. Because service names and granularity may vary across decompositions, the comparison relied on lexical identity and the semantic alignment of responsibilities. While this choice avoids unfair penalization of alternative but reasonable labels, it also introduces interpretive judgment into the matching procedure. We mitigated this threat by applying explicit comparison criteria and documenting the resulting artifacts in the supplementary material.

A third threat concerns the evaluation criteria themselves. Metrics such as precision, recall, and F1-score capture overlap with the reference architecture, but they do not fully reflect architectural soundness. Likewise, expert criteria such as correctness, completeness, modularity, and plausibility improve interpretability, but they still represent a partial view of architectural quality. For this reason, we combined automated comparison with quantitative and qualitative expert assessment.

Finally, the FS condition introduces a specific construct threat. The exemplar came from a different system and did not reproduce the target requirement text, but the study did not formally measure its similarity to Bookstore and PetClinic at the domain or decomposition-pattern levels. It may therefore influence the model beyond clarifying the output format by suggesting a decomposition style, granularity, or interaction density. Different exemplars could produce different outcomes. Moreover, the current design does not determine whether a more detailed natural-language instruction without an explicit example would provide equivalent guidance. A factorial comparison among concise instructions, enriched instruction-only prompts, and multiple exemplars is needed to isolate these effects.

\textbf{Internal Validity}. 

A key internal validity threat is the stochastic variability of LLM outputs. The study used one execution for each system and prompting condition; consequently, it does not quantify variance across repeated runs or establish the stability of the reported decompositions. Keeping the API configuration and prompt templates unchanged and releasing the exact outputs improves procedural reproducibility, but it does not eliminate sampling variability. Prompt phrasing, platform updates, and model-version changes may also alter the results. Future replications should use multiple independent executions per condition and report distributions or confidence intervals for the structural metrics.

Another internal threat concerns the curation of the textual requirements. Although the curation process was intended only to reduce redundancy, ambiguity, and lexical inconsistency, any reformulation of the original requirements may affect what the model perceives as architecturally salient. We mitigated this by preserving the original functionality and business intent of the systems and by using the same curated requirements across prompting conditions.

A further threat concerns expert evaluation. Although the experts had relevant professional experience and reviewed the architectures under blinded and randomized conditions, their judgments may reflect individual preferences, prior architectural beliefs, or different interpretations of the four dimensions. We mitigated ambiguity by providing explicit operational definitions and by retaining individual ratings and divergent comments instead of forcing consensus. Nevertheless, the small expert sample and the descriptive consolidation limit statistical and interpretive generalization.

\textbf{External Validity}. 

This study was conducted on two relatively small systems with well-documented requirements: Bookstore and PetClinic. Their bounded size, limited number of services, and comparatively clear domains are suitable for controlled inspection, but they do not represent the scale, heterogeneity, legacy constraints, or interaction complexity of many industrial systems. The observed results may therefore change for larger architectures, safety-critical domains, noisy or conflicting requirements, and systems with asynchronous, event-driven, or dynamically configured dependencies.

A related external-validity threat concerns model, language, and artifact style. All prompts and curated requirements were written in English, and all generations used OpenAI o3. The findings directly support conclusions about this model in the reported setting, not about every state-of-the-art LLM. Replication should cover additional case studies, substantially larger systems, multilingual and less standardized requirements, multiple model families and versions, and repeated executions within each condition.

\textbf{Reliability}. 
Reliability concerns the replicability of the study procedures and findings. To support reproducibility, we released the curated requirements, reference architectures, prompt templates, generated outputs, evaluation forms, and supporting materials in a public repository~\cite{dataset}. We also documented the main stages of the study, including artifact preparation, generation, normalization, comparison, and expert assessment.

However, some threats remain. Model behavior may change over time due to API, platform, or model-version updates beyond our control; recording the execution period, prompts, and outputs supports auditability but cannot guarantee identical future responses. The synthesis of open-ended comments was author-led and descriptive: responses were organized, grouped by semantic similarity, and compared across alternatives, but no independent coding round or inter-rater agreement statistic was used. Researcher interpretation may therefore affect the resulting categories. Future work should preregister a formal qualitative protocol, employ independent coders, report agreement and disagreement-resolution procedures, and preserve a complete audit trail from quotations to categories.

\section{Final Remarks}
\label{sec:final_remarks}

This study investigated whether an LLM, specifically the OpenAI o3 model, can synthesize microservice architectural descriptions directly from natural-language requirements. To this end, we evaluated the model’s ability to identify service boundaries, responsibilities, and inter-service interactions across two representative software systems under ZS and FS prompting strategies.

For \textbf{RQ1}, the recorded OpenAI o3 outputs recovered most reference services from requirements alone, even in the ZS condition. Across the two systems, ZS correctly identified 11 of the 14 expected services, with a precision, recall, and F1-score of approximately 0.79. Under FS prompting, the model recovered all 14 expected services and introduced only one extra service overall, increasing precision to 0.93, recall to 1.00, and F1-score to approximately 0.97. In this evaluated setting, exemplar-based prompting produced closer agreement with the implemented service boundaries.

For \textbf{RQ2}, the same general pattern emerged, although interaction recovery proved more challenging than service identification. In ZS, the model recovered most of the expected communication links, but at the cost of substantial over-generation, yielding high recall (0.84) but low precision (0.48), with an F1-score of approximately 0.61. Under FS prompting, interaction recovery improved considerably: the model correctly recovered 36 of the 38 expected links, while substantially reducing unsupported connections, resulting in a precision of 0.72, a recall of 0.95, and an F1-score of approximately 0.82. This suggests that FS prompting improves coverage and constrains the tendency of the model to produce overly dense architectural topologies.

For \textbf{RQ3}, expert evaluation showed that the FS architectures were the most positively perceived alternatives across both case studies. Experts rated them highest in correctness, completeness, and plausibility, indicating that they provided the best balance between requirement coverage and architectural organization. In contrast, the ZS architectures were often seen as functionally rich but structurally risky, mainly due to excessive coupling and implausible interaction patterns. The reference architectures were generally perceived as cleaner and more stable, but also less comprehensive in functional coverage. Taken together, these results indicate that architectural usefulness depends on achieving both functional coverage and structurally coherent decomposition.

For \textbf{RQ4}, the qualitative analysis further showed that the limitations of LLM-generated architectures are not random, but concentrated in recurring patterns such as omitted low-salience responsibilities, unsupported infrastructural additions, and overly dense communication structures. These patterns help explain expert preferences and provide useful guidance for future prompt design and architectural review workflows.

Overall, the findings provide bounded empirical evidence that OpenAI o3 can support requirements-driven microservice design as an assistant for early-stage architectural exploration. Its main observed strength lies in identifying candidate services and rapidly proposing alternatives from textual requirements. Interaction design remains more error-prone and requires expert validation of coupling, cross-cutting concerns, and deployment plausibility. The single FS exemplar was associated with closer reference agreement and more favorable expert perceptions, but the present design cannot separate the effects of output-format demonstration, exemplar-specific structural guidance, and stochastic variation. The results therefore motivate decision-support use and further controlled evaluation, not replacement of expert-driven design.

Future work should expand this investigation in at least five directions. First, replications should include substantially larger and architecturally more complex systems from additional domains, including industrial and safety-critical contexts. Second, multiple LLM families and model versions should be compared, with repeated independent executions per condition to quantify output variability and metric stability. Third, controlled studies should disentangle exemplar effects by comparing concise ZS prompts, richer natural-language instruction-only prompts, and multiple exemplars that vary in domain and decomposition pattern. Fourth, longitudinal human-in-the-loop studies should examine how architects revise generated decompositions and whether the suggestions improve design outcomes over time. Finally, requirement-driven synthesis should be combined with static, dynamic, and repository-based architectural evidence so that generated alternatives can be validated against implementation constraints rather than evaluated from requirements alone.

\section*{ARTIFACT AVAILABILITY}

All artifacts used in this study — including requirements, prompting strategies, architectural outputs, evaluation scripts, and expert review templates — are available to ensure transparency, facilitate replication, and support future research endeavors~\cite{dataset}.

\section*{ACKNOWLEDGMENTS}
This work has been partially funded by the project ``iSOP Base: Investigação e desenvolvimento de base arquitetural e tecnológica da Intelligent Sensing Operating Platform (iSOP)'' supported by CENTRO DE COMPETÊNCIA EMBRAPII VIRTUS EM HARDWARE INTELIGENTE PARA INDÚSTRIA - VIRTUS-CC, with financial resources from the PPI HardwareBR of the MCTI grant number 055/2023, signed with EMBRAPII.

\renewcommand\refname{REFERENCES}
\bibliographystyle{ACM-Reference-Format}
\bibliography{sn-bibliography}


\end{document}